\documentclass[a4paper,fleqn,usenatbib,useAMS]{mnras}

\usepackage{graphicx}	
\usepackage{graphics}
\usepackage{amsmath}	
\usepackage{amssymb}	
\usepackage{amsfonts}
\usepackage{multicol}        
\usepackage{bm}		
\usepackage{pdflscape}	
\usepackage{natbib}
\usepackage{epstopdf}

\newcommand{\ellb }{\boldsymbol{\ell }}

\newcommand{\Lb }{\boldsymbol{L }}

\def\bdw{\boldsymbol{w }}
\def\bda{\boldsymbol{a }}
\def\bdb{\boldsymbol{b }}

\usepackage[T1]{fontenc}
\usepackage{ae,aecompl}

\title[Impact of SZ cluster residuals in CMB maps]{Impact of SZ cluster residuals in CMB maps and CMB-LSS cross-correlations}

\author[T. Chen, M. Remazeilles, C. Dickinson]{T.\,Chen,\!\thanks{E-mail:~\url{tianyue.chen@manchester.ac.uk}} M.\,Remazeilles,\!\thanks{E-mail:~\url{mathieu.remazeilles@manchester.ac.uk}} C.\,Dickinson
\\
Jodrell Bank Centre for Astrophysics, Alan Turing Building, School of Physics and Astronomy, The University of Manchester, \\
Oxford Road, Manchester, M13 9PL\\}

\begin{document}
\label{firstpage}
\pagerange{\pageref{firstpage}--\pageref{lastpage}}
\maketitle


\begin{abstract}
Residual foreground contamination in cosmic microwave background (CMB) maps, such as the residual contamination from thermal Sunyaev-Zeldovich (SZ) effect in the direction of galaxy clusters, can bias the cross-correlation measurements between CMB and large-scale structure optical surveys. It is thus essential to quantify those residuals and, if possible, to null out SZ cluster residuals in CMB maps. We quantify for the first time the amount of SZ cluster contamination in the released \emph{Planck} 2015 CMB maps through (i) the stacking of CMB maps in the direction of the clusters, and (ii) the computation of cross-correlation power spectra between CMB maps and the SDSS-IV large-scale structure data.  Our cross-power spectrum analysis yields a $30\sigma$ detection at the cluster scale ($\ell=1500$--2500) and a $39\sigma$ detection on larger scales ($\ell=500$--1500) due to clustering of SZ clusters, giving an overall $54\sigma$ detection of SZ cluster residuals in the \emph{Planck} CMB maps. The \emph{Planck} 2015 NILC CMB map is shown to have $44\pm4\%$ of thermal SZ foreground emission left in it. Using the 'Constrained ILC' component separation technique, we construct an alternative \emph{Planck} CMB map, the 2D-ILC map, which is shown to have negligible SZ contamination, at the cost of being slightly more contaminated by Galactic foregrounds and noise. We also discuss the impact of the SZ residuals in CMB maps on the measurement of the ISW effect, which is shown to be negligible based on our analysis.
  
\end{abstract}




\section{Introduction}

The interactions between the large-scale structures (LSS) and the cosmic microwave background (CMB) radiation in the Universe lead to gravitational and spectral distortions to the CMB radiation. These include weak gravitational lensing (WGL) effects on the CMB \citep{Lewis2006}, deflection of the background CMB light by LSS; Sunyaev-Zeldovich (SZ) effect \citep{Sunyaev1972} due to the hot electron gas in galaxy clusters scattering off the CMB radiation, and the integrated Sachs-Wolfe (ISW) effect \citep{Sachs1967,Rees1968}, a blue shift of the CMB photons when they travel through LSS gravitational potential wells, which are diluted forward in time by the accelerated expansion of the Universe. Studying the signature of these CMB distortions enables to probe dark matter and dark energy. The released \emph{Planck} maps of the CMB lensing field \citep{Planck_2015_lensing}, thermal SZ $y$-Compton parameter \citep{planck2015sz}, and ISW effect \citep{planckXXI}, all provide \emph{indirect} tracers of the matter distribution in the sky through the imprint of the large-scale structures on the CMB. Conversely, optical surveys of large-scale structures, such as BOSS/SDSS-III \citep{BOSS2013} and future surveys by LSST \citep{LSST2009} and \emph{Euclid} \citep{Euclid2011}, offer \emph{direct} tracers of the matter distribution in the Universe. 

In the cosmology community, there is a growing interest in cross-correlating multiple tracers of dark matter issued from complementary data sets, such as CMB and LSS surveys. \citealt{gvh+18} have cross-correlated the \emph{Planck} CMB lensing map with the SDSS galaxy density map to measure the scale-dependence of the galaxy bias and constrain neutrino masses. Cross-power spectra between \emph{Planck} CMB lensing and galaxy lensing shear maps have been computed to measure the galaxy lensing shear bias \citep{Liu2015} and the amplitude of the gravitational lensing effect \citep{Singh2017}. The cross-correlations between CMB maps and radio source surveys \citep[e.g.,][]{bc02,bc04,gcn+12}, infrared surveys \citep[e.g.,][]{als04,gcn+12,gsg12, sw16}, X-ray background \citep[e.g.,][]{bc04,gcn+12}, and optical surveys \citep[e.g.,][]{phs+05,gcn+06,planckXIX,planckXXI, bpf+17}, have been used to measure the ISW effect and constrain dark energy. The cross-correlations between CMB maps and SDSS data \citep[e.g.,][]{ps00,Sherwin2012}, 2MASS data \citep{als04}, and \emph{Planck} thermal SZ map \citep{hs14} have been used to constrain cosmological parameters such as the equation of state of dark energy, $w$, and the r.m.s fluctuations of dark matter, $\sigma_8$. \cite{mvh+15} and \cite{hmh+15} constrained cosmological parameters, hydrostatic mass bias, and baryon component, by cross-correlating the CFHTLenS mass map with the \emph{Planck} thermal SZ map. The cross-correlations between independent tracers of dark matter from CMB and LSS experiments has the advantage of mitigating instrumental systematics, which will likely be uncorrelated between the datasets. They also provide a way to calibrate the ``mass biases'' \citep{Sherwin2012,Das2013,als+15}, i.e., the ratio between the observable mass distribution and the underlying dark matter mass distribution (e.g., the galaxy bias from LRG or radio galaxies), therefore allowing a more robust measurement of the matter power spectrum. 

However, CMB-LSS cross-correlations may suffer from another kind of systematic error: the contamination of the CMB products by LSS foregrounds, such as the residual emission from SZ galaxy clusters in the CMB maps. As mentioned in \cite{dpv+17}, when performing CMB-LSS cross-correlations, a possible source of systematic error is due to the fact that the residual SZ clusters in the CMB maps must host some of the galaxies in the LSS survey. In \cite{gvh+18}, the scale-dependent galaxy bias measured through CMB lensing-SDSS cross-correlation was found to be lower than expected, which was attributed to possible thermal SZ contamination in the \emph{Planck} CMB lensing map. Also, \cite{mh18} have shown that the SZ contamination of the CMB lensing map causes significant bias on the dark matter halo masses when measured from cross-correlations between CMB lensing and galaxy density fields. 

LSS residuals in the CMB products can thus no longer be ignored in the context CMB-LSS cross-correlations for unbiased measurements. Although the contamination by LSS foregrounds has been mitigated in the public CMB maps through component separation algorithms, there is always a non-zero residual of thermal SZ effect from galaxy clusters \citep[see e.g.,][]{Planck17-dispersion}. \cite{lgm+15} have also observed SZ residuals by stacking the \emph{Planck} 2013 SMICA, NILC and SEVEM CMB temperature maps in the direction of galaxy clusters. In particular, SZ residuals in CMB maps may lead to spurious, non-physical, (anti)-correlations with galaxy surveys. Those SZ residuals must also propagate to the CMB lensing potential map in a non-trivial way since it is derived from the CMB map itself by means of quadratic estimators \citep[e.g.,][]{Hu2002}. Therefore, for unbiased CMB-LSS cross-correlations, it is essential to perfectly control the LSS contamination in CMB maps. 

CMB foreground residuals can impact both ISW and lensing measurements, and as a consequence might lead to confusion in constraints on dark energy and neutrino masses. Indeed, thermal SZ residuals appear as negative temperature fluctuations (blue) in the CMB maps because most of the component separation weights for CMB reconstruction actually come from the 100--143\,GHz frequency channels, where the CMB is dominant but the thermal SZ spectral energy distribution (SED) is negative. Negative SZ residuals in the CMB will thus anti-correlate with galaxy flux/number density from LSS surveys, therefore causing a deficit of power on small scales in the matter cross-power spectrum, which might be confused with effects from massive neutrinos.   

In this work we show that there is indeed a detectable SZ signature  left in current {\it Planck} CMB maps by correlation with SDSS-IV LSS data. We quantify the amount of spurious correlation between SZ cluster residuals in the released \emph{Planck} CMB maps and the SDSS-IV galaxy survey. We also propose an alternative \emph{Planck} CMB map, which is free from SZ residuals, that we recommend for cross-correlation studies between CMB and LSS surveys.

The paper is organised as follows.  Sect.\,\ref{datasec} describes the data sets used in our analysis.  Sect.\,\ref{anasec} gives the first sight of the correlation between the SZ cluster residuals and the LSS through stacking analysis.  Sect.\,\ref{cross_spec} presents the results from the cross-power spectrum analysis and discusses the effect of the SZ cluster residuals on the ISW detection.  Sect.\,\ref{dissec} summarizes the results and draws the conclusions. 


\section{Data description}\label{datasec}

In this section, we describe the set of CMB temperature maps (Sect.\,\ref{secnilc} and \ref{sec2dilc}) and the LSS optical survey map (Sect.\,\ref{bossdata}) that we use for the cross-correlation analysis. A thermal SZ map at 143\,GHz (Sect.\,\ref{szcat}) is also generated from the \emph{Planck} 2015 SZ catalogue \citep{planckXXVII} for simulations.  

\subsection{The \textit{Planck} \textsc{NILC} CMB map} \label{secnilc}

Four CMB maps have been released by \emph{Planck} \citep{Planck_2015_CMB}. Each of them has been estimated by an independent component separation algorithm: \textsc{Commander} \citep{Eriksen2008}, a Bayesian parametric pixel-by-pixel fitting with MCMC Gibbs sampling; \textsc{SEVEM} \citep{Fernandez2012}, an internal template fitting in wavelet space; \textsc{SMICA} \citep{Cardoso2008}, a power-spectra fitting approach in harmonic space; \textsc{NILC} \citep{Delabrouille2009}, a minimum-variance internal linear combination in needlet space. The four \emph{Planck} CMB products have shown good consistency at a level of a few $\mu$K \citep{Planck_2015_CMB}, with minimized residuals from foreground contamination (Galactic foreground emissions, SZ, extragalactic sources). The \emph{Planck} CMB maps have been mapped on the sphere through a \textsc{healpix} pixelization scheme (\citealt{ghb+05}) on a $N_{\rm side}=2048$ grid (pixel size $\approx 1.7$\,arcmin). The beam resolution of the \emph{Planck} CMB maps is $5$\,arcmin.

While we consider the four \emph{Planck} CMB maps for our stacking analysis in Sect.~\ref{stacking}, we focus on the \emph{Planck} \textsc{NILC} CMB map for the CMB-LSS cross-power spectrum analysis in Sect.~\ref{cross_spec}. 

The basics of the \textsc{NILC} component separation algorithm can be summarized as follows. In intensity units, the data, $x_\nu$, in each frequency band, $\nu$, and in each pixel are the superposition of the CMB temperature fluctuations, $s\equiv \Delta T$, and the contamination, $n_\nu$, which include foregrounds and noise:
\begin{equation}
x_\nu = a_\nu\, s\, +\, n_\nu,
\label{yinzhe}
\end{equation}
where $a_\nu\equiv dB_\nu(T)/dT\vert_{T=T_{\rm CMB}}$ is the derivative of the blackbody spectrum $B_\nu(T)$ with respect to temperature $T$, i.e. the spectral energy distribution (SED) of the CMB temperature anisotropies, $s$, across frequencies. The \textsc{NILC} method then consists of estimating the CMB signal, $\widehat{s}$, as a weighted internal linear combination (ILC) of the \emph{Planck} frequency maps $x_\nu$,
\begin{equation}
\widehat{s} = \sum_\nu\, w_\nu\, x_\nu,
\end{equation}
that is constrained to give unit response to the CMB SED, i.e.,
\begin{equation}
\label{sed1}
\sum_\nu\, w_\nu\, a_\nu = 1,
\end{equation}
and to be of minimum variance, i.e.,
\begin{equation}
\label{minvar}
{\partial\, \langle\, \widehat{s}\,^2\, \rangle \over \partial\, w_\nu}\, \equiv\, {\partial\over\partial\, w_\nu}\, \left(\sum_{\nu'}\,\sum_{\nu''}\, w_{\nu'}\,\mathrm{C}_{\nu'\nu''}\,w_{\nu''}\right)\, =\, 0,
\end{equation}
where matrix $\mathrm{C}_{\nu\nu'}\equiv\langle\,x_\nu x_{\nu'}\,\rangle$ is the frequency-by-frequency covariance matrix of the data ($9\times 9\times N$ matrix, where $9$ is the number of \emph{Planck} frequencies and $N$ is the total number of pixels in \emph{Planck} maps). The elements of the covariance matrix are computed in each pixel $p$ as
\begin{equation}
\label{covar}
\mathrm{C}_{\nu\nu'}(p) = {1\over N_p}\sum_{p' \in \mathcal{D}(p)} x_\nu(p')x_{\nu'}(p'),
\end{equation}
where the sum runs over $N_p$ pixels $p'$ of a circular domain $\mathcal{D}(p)$ surrounding the pixel $p$. The choices for the size and morphology of the pixel domains $\mathcal{D}(p)$ have been described in \cite{Basak2012}.
Using a Lagrange multiplier, the solution for the vector of NILC weights, $\bdw=\{w_\nu\},$ is thus given by 
\begin{equation}
\label{nilc0}
\bdw^t = {\bda^t\boldsymbol{\mathrm{C}}^{-1}\over\bda^t\,\boldsymbol{\mathrm{C}}^{-1}\,\bda},
\end{equation}
where the upperscript $t$ stands for transposition.
The \textsc{NILC} estimate is then
\begin{equation}
\label{nilc}
\widehat{s} = s\, +\, \sum_\nu\, w_\nu\, n_\nu,
\end{equation}
thus providing an unbiased estimate of the CMB, $s$, thanks to the constraint in Eq.~\ref{sed1}, while residual foregrounds, $\sum_\nu w_\nu n_\nu$, are minimized thanks to the condition in Eq.~\ref{minvar}. The \textsc{NILC} reconstruction of the CMB is performed in needlet (spherical wavelet) space \citep{Narcowich2006,Guilloux2007} to allow the ILC weights, $w_\nu$, to vary over the sky, adjusting to the local conditions of contamination both over the sky and angular scale.


\subsection{The \textsc{2D-ILC} CMB map}\label{sec2dilc}

The four component separation pipelines applied to the \emph{Planck} data have been optimized to minimize the global contamination from astrophysical foregrounds and instrumental noise. However, a certain amount of residual foregrounds is inevitably present in those CMB maps, at different levels depending on the area of the sky and the angular scale. 

Thermal SZ residuals will be left in the \textsc{NILC} CMB map (Eq.~\ref{nilc}) through the following expression:
\begin{equation}
\label{szres}
\sum_\nu\, w_\nu\, b_\nu\, y,
\end{equation}
where $w_\nu$ are the NILC weights, $b_\nu$ is the SED of the thermal SZ effect:
\begin{equation}
b_\nu = x \coth(x/2)-4\,,\mbox{with }\, x \equiv h\nu/(k_{\text{B}}T_{\text{CMB}})\,, 
\end{equation}
and $y$ is the SZ $y$-Compton parameter. Since the signal-to-noise ratio is favourable to the CMB in the $100$--$143$\,GHz frequency range, the bulk of the \textsc{NILC} weights is attributed mostly to this frequency range, for which the thermal SZ SED is negative relative to the CMB. As a consequence, thermal SZ residuals in CMB maps appears as negative fluctuations in the direction of the galaxy clusters (see Fig.~\ref{fig:planckstack}). 

Depending on the scientific objective, in particular cross-correlations between CMB and galaxy surveys, it might be more useful to \emph{nullify} specific LSS foregrounds in the CMB map, such as the thermal SZ contamination from galaxy clusters, rather than \emph{minimizing} the global contamination.

In this context, we also use in this work the \textsc{2D-ILC} CMB map \citep[e.g.,][]{Planck17-dispersion}, a thermal SZ-free CMB map which we have produced by applying the 'Constrained ILC' component separation method \citep{Remazeilles2011a} to the \emph{Planck} 2015 data. The Constrained ILC method is similar to \textsc{NILC} as it is a weighted linear combination of the frequency maps in needlet space that offers unit response to the CMB spectrum, but has an additional constraint of giving zero response to the thermal SZ spectrum. In other words, the vector of weights, $w_\nu$, for the Constrained ILC is constructed to be orthogonal to the thermal SZ SED vector, $b_\nu$, so that Eq.~\ref{sed1} is replaced by  
\begin{subequations}
\begin{align}
\label{sed2-1}
\sum_\nu\, w_\nu\, a_\nu = 1,
\\
\label{sed2-2}
\sum_\nu\, w_\nu\, b_\nu = 0.
\end{align}
\end{subequations}
In this way, the thermal SZ contamination Eq.~\ref{szres} is nulled out, thus guaranteeing the total absence of thermal SZ residuals in the resulting \textsc{2D-ILC} CMB map\footnote{The name \textsc{2D-ILC} comes from the two-dimensional (2D) constraint in Eqs.~\ref{sed2-1}-\ref{sed2-2}.}. A unique solution to Eqs.~\ref{minvar} and \ref{sed2-1}-\ref{sed2-2} is derived by using Lagrange multipliers \citep{Remazeilles2011a}:
\begin{align}
\label{2d-ilc-weights}
\bdw^t = { \left( \bdb^{t} \boldsymbol{\mathrm{C}}^{-1} \bdb \right)
  \bda^{t} {\boldsymbol{\mathrm{C}}}^{-1} - \left( \bda^{t}
  \boldsymbol{\mathrm{C}}^{-1} \bdb \right) \bdb^{t}
  {\boldsymbol{\mathrm{C}}}^{-1}
  \over
  \left(\bda^{t} \boldsymbol{\mathrm{C}}^{-1} \bda
  \right)\left(\bdb^{t} \boldsymbol{\mathrm{C}}^{-1} \bdb \right) -
  \left( \bda^{t} \boldsymbol{\mathrm{C}}^{-1} \bdb \right)^2 }
\end{align}

The SZ-free \textsc{2D-ILC} CMB map\footnote{The \textsc{2D-ILC} used in this work is available upon request to the authors.} has been produced at the same \textsc{healpix} $N_{\rm side} = 2048$ resolution and $5$\,arcmin beam size than the public \emph{Planck} CMB maps by assigning to the nine \emph{Planck} frequency maps the specific weights Eq.~\ref{2d-ilc-weights} that fulfill the constraints of Eqs.~\ref{sed2-1}-\ref{sed2-2}.

The Constrained ILC method (and 2D-ILC map) is the first solution proposed in the literature \citep{Remazeilles2011a} to null out thermal SZ residuals in CMB maps, and its unique property was used to detect the kinetic SZ effect in the \emph{Planck} data by \cite{planck2013-XIII}. An alternative approach based on sparsity (LGMCA) has then been proposed by \cite{Bobin2014,Bobin2016} to achieve the same goal as the 2D-ILC map. The LGMCA CMB map has also been used to measure the kinetic SZ effect by \cite{Hill2016,Ferraro2016}. While the LGMCA CMB map has been produced from the combination of \emph{Planck} and \emph{WMAP} data, the 2D-ILC CMB map, like the released Planck 2015 CMB maps, is based on internal \emph{Planck} data. Therefore for internal consistency with the released Planck 2015 CMB maps we use the 2D-ILC map as SZ-free CMB template in most of our analysis, and we consider the alternative LGMCA map in Sect.~\ref{sec-cmbxsdss} for consistency check of our results.


\subsection{SDSS catalogue}\label{bossdata}

The LSS dataset used in this paper is the main photometric galaxy sample from the DR13 release of the SDSS-IV survey \citep[hereafter MphG,][]{aaa+17}. The MphG catalogue was downloaded from the SDSS DR13 database\footnote{http://www.sdss.org/dr13/}. The total number of galaxies in this release is about 208 millions, covering a sky area of 14555 deg$^2$ \citep{aaa+17}. 

SDSS probes galaxies within five optical filter bands $u, g, r, i, z$.  Five different measurements of the magnitude, derived using different fitting methods in the SDSS pipeline, are given in each band for each source. The composite model (hereafter cModel) flux is the combined flux of the best-fitting exponential and de Vaucouleurs fluxes in each band, where the fraction of each term is optimised to give the best-fitting to the source profile \citep{slb+02}. The cModel magnitude is used in our analysis since it is an adequate proxy to use as a universal magnitude for all types of objects and a reliable estimate of the galaxy flux under most conditions\footnote{http://www.sdss.org/dr12/algorithms/magnitudes} \citep{slb+02}. 

We use the $r$ band for our analysis because this band has a better sensitivity \citep[$r<22.2$ for $95\%$ completeness for point sources,][]{yaa+00} and photometric calibration accuracy \citep[$0.8\%$,][]{psf+08}. The $r$ band is also least affected by dust extinction.  We discard faint sources with the $r$ band cModel magnitude below the completeness level $r<22.2$ as those faint sources can smear the cross-correlation signal by adding uncorrelated background noise.  We also discard the brightest sources ($r<17$ amounting to $\approx1\%$ ) to avoid a small number of bright sources dominating the statistical results. After the selection, the number of galaxies in our sub-sample is about 133 million, which is $\sim 64\%$ of the galaxies in the SDSS DR13 MphG sample. We emphasise that the main results are not strongly dependent on the exact choice of catalogue. 

The galaxies detected by SDSS have a typical size of a few arcsec \citep[e.g.,][]{aaa+17,slb+02}, much smaller than the $\sim 1.7$\,arcmin pixel size of the \emph{Planck} maps having a \textsc{healpix} $N_{\rm side} = 2048$. Therefore, each selected SDSS source is assigned into a single \textsc{healpix} $N_{\rm side} = 2048$ pixel corresponding to its sky coordinates given by the MphG catalogue. A density contrast map is then constructed from the selected sources, where the density contrast is defined as
\begin{equation}
n = \frac{(N-\bar{N})}{\bar{N}}\,,
\end{equation}
with $N$ being the number of SDSS sources in each pixel and $\bar{N}$ the average number of sources per pixel. The density contrast map is then convolved with a $5$\,arcmin Gaussian beam using the \textsc{healpix} \textsc{smoothing} routine (\citealt{ghb+05}) in order to have consistent beam resolution with the \emph{Planck} CMB maps described in Sect.\,\ref{secnilc} and \ref{sec2dilc}.  


\subsection{SZ catalogue}\label{szcat}

In order to interpret the structure of the spurious correlation signal between CMB and galaxy surveys due to large-scale structure residuals in CMB maps, we also construct a pure thermal SZ map at 143\,GHz (hereafter, catalogue SZ map) from the \emph{Planck} 2015 SZ catalogue (\citealt{planckXXVII}). The total number of galaxy clusters in the \emph{Planck} 2015 SZ catalogue is 1653. For each galaxy cluster, the \emph{Planck} catalogue provides the value of the integrated thermal SZ flux $Y_{5R500}$:
\begin{equation}\label{equy5r500}
Y_{5R500} = \hat y_0 \int_{\theta < 5\times \theta_{500}}dr\tau_{\theta_s}(r)\,,
\end{equation}
where $\hat y_0$ is the Comptonization parameter, $\tau_{\theta_s}$ is the thermal SZ pressure profile of the cluster, and $\theta_{500}$ is the radius within which the average density is 500 times the critical density of the Universe, and is related to the cluster angular size, $\theta_s$, by $\theta_{500} = c_{500}\times \theta_s$, with $c_{500}  = 1.177$ (\citealt{planckXXVII}). The cluster angular size $\theta_s$ and integrated flux $Y_{5R500}$ are given in the form of a joint probability distribution in the ($\theta_s$, $Y_{5R500}$) plane.  For each SZ cluster, the integrated thermal SZ flux $Y_{5R500}$ and the cluster angular size $\theta_s$ are determined from the maximum likelihood in the ($\theta_s$,\,$Y_{5R500}$) plane. We use a minimal value  of 1.7\,arcmin for $\theta_s$, corresponding to the pixel size. Since our cross-power spectrum analysis is not sensitive to the exact profile of the SZ clusters, for simplicity we assume each SZ cluster to have a circular top-hat profile with radius $\theta_s$ and uniform brightness. Under this assumption, Eq.~\ref{equy5r500} reduces to 
\begin{equation}\label{y5r500}
Y_{5R500} = \hat y_0\times\pi\times\theta_{s}^2\,,
\end{equation}
from which we derive the Compton parameter, $\hat y_0$, for each cluster. Since the majority of the clusters have $\theta_s<$5\,arcmin, and the beam size is 5\,arcmin, this assumption is valid.

For a power spectrum analysis, one has to be careful that the spectrum is not dominated by a small number of bright sources, which can cause a bias relative to the average distribution of sources. We confirmed that this is not the case in our analysis by removing the brightest sources and repeating the analysis. We found that excluding the brightest 1--10\,\% of clusters made no significant difference to the results. When excluding $>10\,\%$ the cross-correlation signal begins to reduce due to the lack of signal. The SZ clusters are projected onto a \textsc{healpix} map of $N_{\rm side} = 2048$, according to their sky coordinates given by the \emph{Planck} SZ catalogue. The pixels included in the circles of radius $\theta_s$ for each cluster are equally given the value $\hat y_0$ of that source.  In each pixel, the values $\hat y_0$ of all SZ clusters falling in that pixel are summed. The catalogue SZ $y$-map is then converted to thermodynamic temperature units at 143\,GHz through the spectral energy distribution (SED) of the thermal SZ effect: 
\begin{equation}\label{tsz}
\frac{\Delta T}{T_{\text{CMB}}} = g(\nu = 143\,{\rm GHz})\,\hat y_0,
\end{equation}
where the non-relativistic thermal SZ SED is given by ${g(\nu) = x \coth(x/2)-4}$, with ${x = h\nu/(k_{\text{B}}T_{\text{CMB}})}$ (\citealt{Sunyaev1972}). At $143$\,GHz, the temperature of the SZ clusters is negative. The choice of $143$\,GHz is determined by the fact that most of residual SZ contamination in \emph{Planck} CMB maps comes from this frequency channel, where the signal-to-noise is favourable to the CMB. Finally, the catalogue map at 143\,GHz is convolved with a $5$\,arcmin Gaussian beam to be consistent with the resolution of the \emph{Planck} CMB and SDSS MphG maps described earlier.


\begin{figure*}
\includegraphics[width=1.0\hsize]{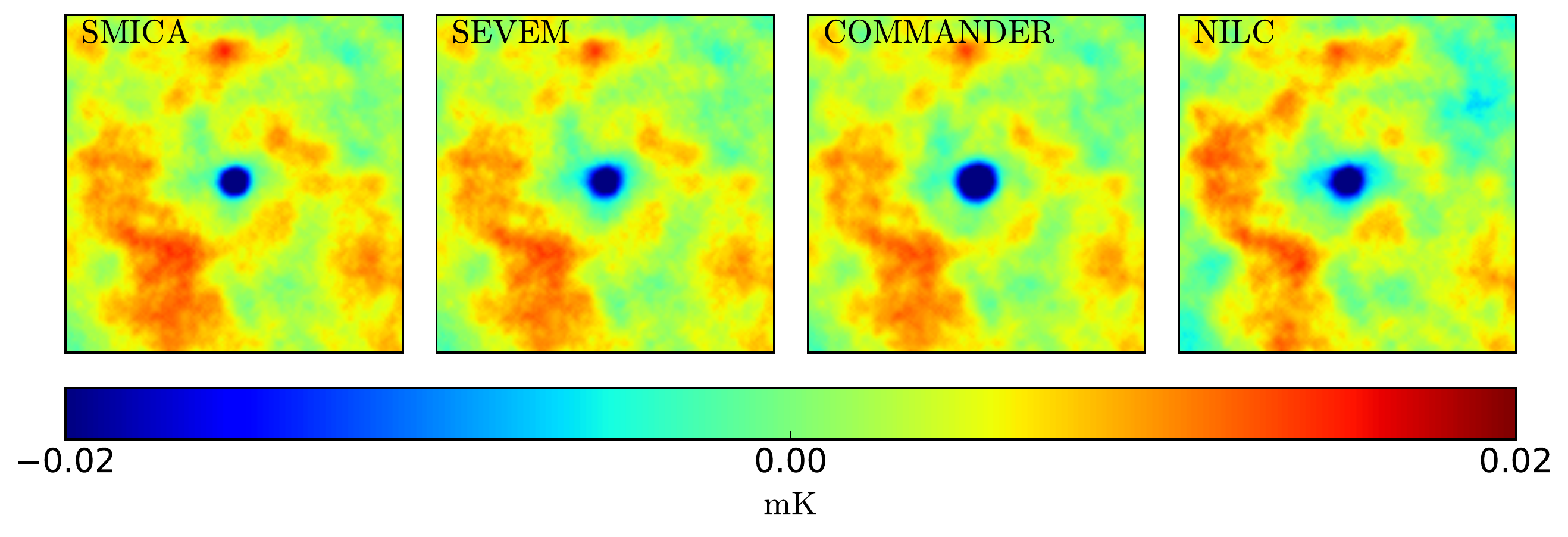}
\caption{Stacked {\it Planck} CMB maps in the directions of the SZ clusters of the \emph{Planck} 2015 SZ catalogue \protect\citep{planckXXVII} in the Galactic coordinate system with the north Galactic pole upwards. From {\it left} to {\it right}: SMICA, SEVEM, COMMANDER and  NILC CMB maps. Each map covers $3^{\circ}\times3^{\circ}$ and the angular resolution is 5\,arcmin. Blue spots in the centre show evidence for negative thermal SZ contamination from galaxy clusters.}
\label{fig:planckstack}
\end{figure*}
\begin{figure*}
\includegraphics[width=1.0\hsize]{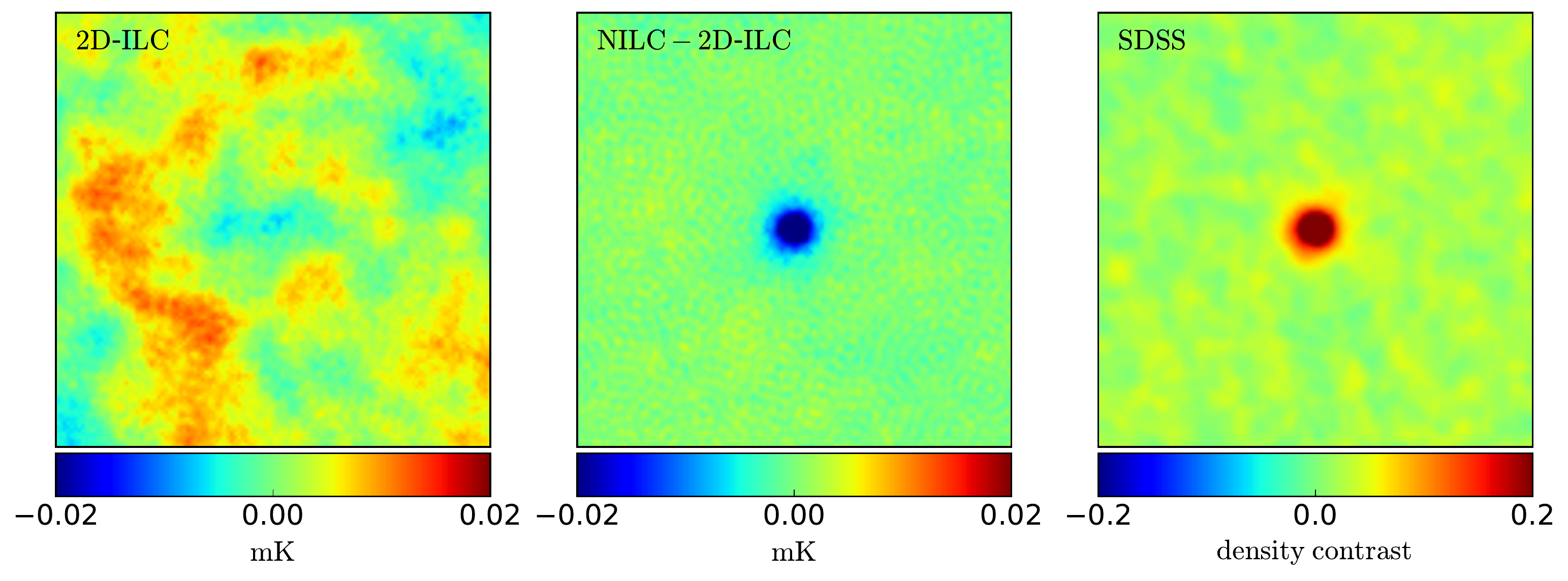}
\caption{Stacked maps in the directions of the SZ clusters of the \emph{Planck} 2015 SZ catalogue \protect\citep{planckXXVII} in the Galactic coordinate system with the north Galactic pole upwards: the 2D-ILC CMB map (\textit{left}), the difference (NILC$-$2D-ILC) map (\textit{middle}), and the SDSS MphG galaxy density map (\textit{right}). Each map covers $3^{\circ}\times3^{\circ}$ and the angular resolution is 5\,arcmin. }
\label{fig:bossstack}
\end{figure*}

\section{Map analysis}\label{anasec}

We first perform visual inspection of the residual SZ contamination in the \emph{Planck} 2015 CMB maps, and highlight the resulting spurious correlation with large-scale structure data.


\subsection{Stacking CMB maps in the direction of galaxy clusters}
\label{stacking}

To demonstrate that there are LSS-correlated residuals in the CMB maps, each of the four \emph{Planck} 2015 CMB maps -- \textsc{SMICA}, \textsc{SEVEM}, \textsc{COMMANDER}, \textsc{NILC} -- is stacked at the locations of the SZ clusters of the \emph{Planck} 2015 SZ catalogue (\citealt{planckXXVII}). Each CMB map is thus projected onto $3^{\circ}\times3^{\circ}$ patches of the sky centred at each SZ cluster location. The patches are then averaged all together, giving the stacked maps shown in Fig.~\ref{fig:planckstack}. Clearly, the four stacked \emph{Planck} CMB maps show strong negative temperature fluctuations in the centre (blue spot), showing clear residual contamination from thermal SZ effect in the direction of galaxy clusters. While four different component separation techniques have been operated on the \emph{Planck} frequency data to minimize the overall foreground contamination and extract the CMB signal, there is still significant residual thermal SZ emission from galaxy clusters in the \emph{Planck} CMB maps. Thermal SZ residuals appear as negative temperature fluctuations in each \emph{Planck} CMB map because the bulk of the weights assigned to the \emph{Planck} frequency maps by component separation algorithms is around $100$--$143$\, GHz frequencies, where the CMB emission is dominant over the foreground emission but the thermal SZ SED is negative.

Following the same procedure, we show in the right panel of Fig.~\ref{fig:bossstack} the SDSS MphG galaxy survey map stacked in the directions of the same clusters from the \emph{Planck} 2015 SZ catalogue. As expected from the agglomeration of galaxies within clusters, the stacked SDSS MphG map shows a strong positive overdensity in the direction of SZ clusters. Our stacking analysis thus provides visual evidence for anti-correlation between SZ cluster residual fluctuations in CMB maps and galaxy overdensities in the SDSS survey. The presence of clusters in the CMB maps may bias any cross-correlation analysis between CMB products and optical galaxy surveys. Therefore, we warn that any statistical interpretation of cross-correlation results (e.g., CMB lensing-galaxy lensing correlations) must be done by bearing in mind the amount of spurious correlations from cluster residuals in CMB maps. 

Depending on the scientific purpose, for example CMB-LSS cross-correlations, it might be more useful to filter out LSS residuals, such as thermal SZ emission from galaxy clusters, in the CMB maps rather than minimizing the global foreground contamination. In this regard, we propose an \mbox{SZ-free} CMB map, termed as 2D-ILC map, which we have constructed from the \emph{Planck} 2015 data using the Constrained ILC component separation technique. The Constrained ILC is specifically designed to null out thermal SZ effects in the CMB map (Sect.~\ref{sec2dilc}). 

The result of stacking the 2D-ILC CMB map in the direction of the clusters of the \emph{Planck} SZ catalogue is shown on the left panel of Fig.~\ref{fig:bossstack}. In the case of the 2D-ILC CMB map there is a clear absence of thermal SZ contamination from galaxy clusters, contrasting against the other \emph{Planck} CMB maps. The cost of this extra filtering constraint is that the 2D-ILC map is slightly noisier (small-scale granularity) than the public \emph{Planck} CMB maps. In the middle panel of Fig.~\ref{fig:bossstack}, we show the result of stacking the difference map between the NILC and 2D-ILC maps. The difference highlights the negative SZ cluster residuals from the NILC CMB map, dominating over the noise fluctuations from the 2D-ILC map, and anti-correlating with the galaxy density contrast from SDSS (right panel). The absence of LSS residuals (or negligible SZ contamination with respect to noise) in the 2D-ILC map makes it particularly suited for cross-correlations studies with LSS optical surveys. 

\begin{figure*}
\includegraphics[width=0.98\hsize]{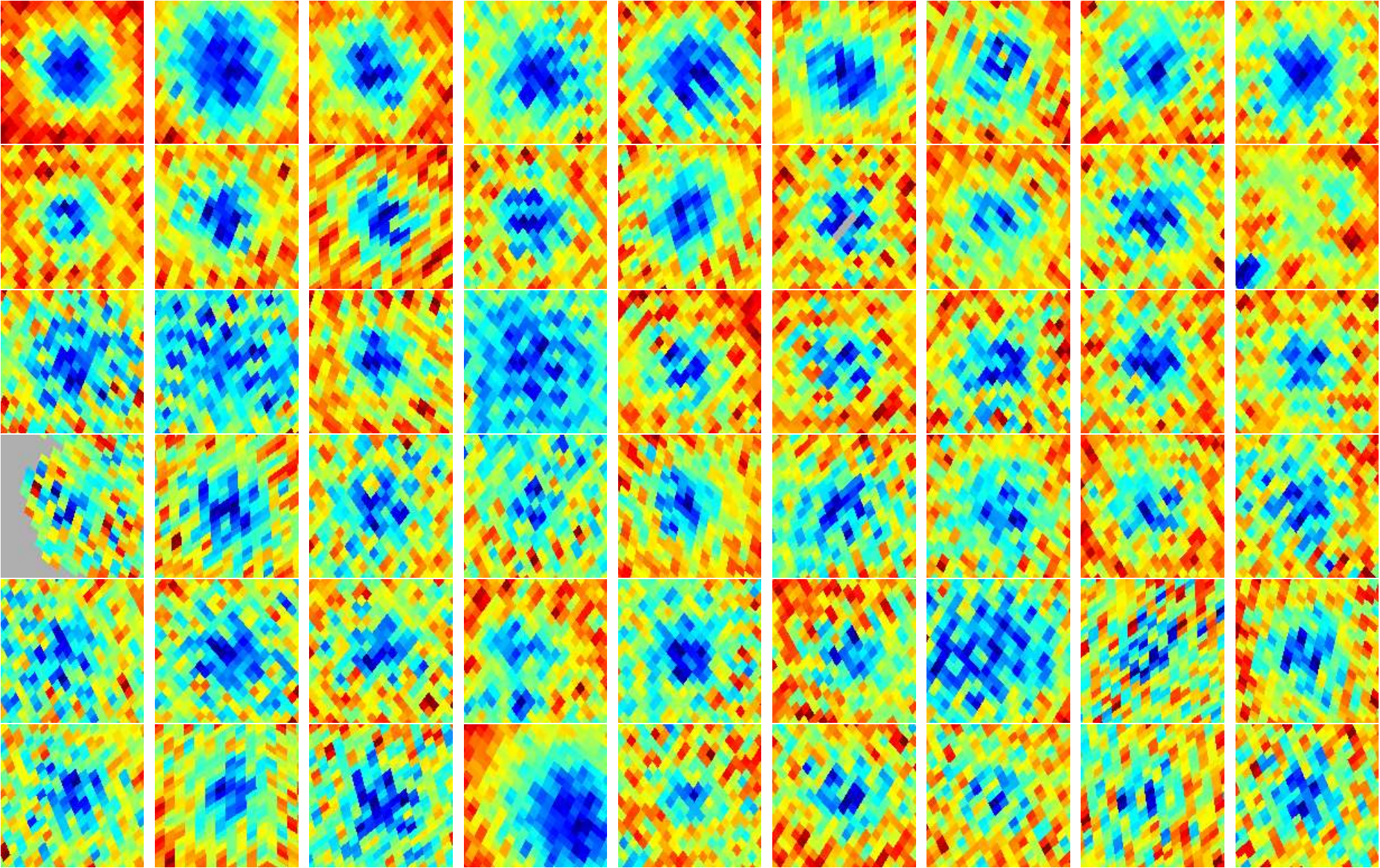}
\hspace*{3pt}\includegraphics[width=0.98\hsize]{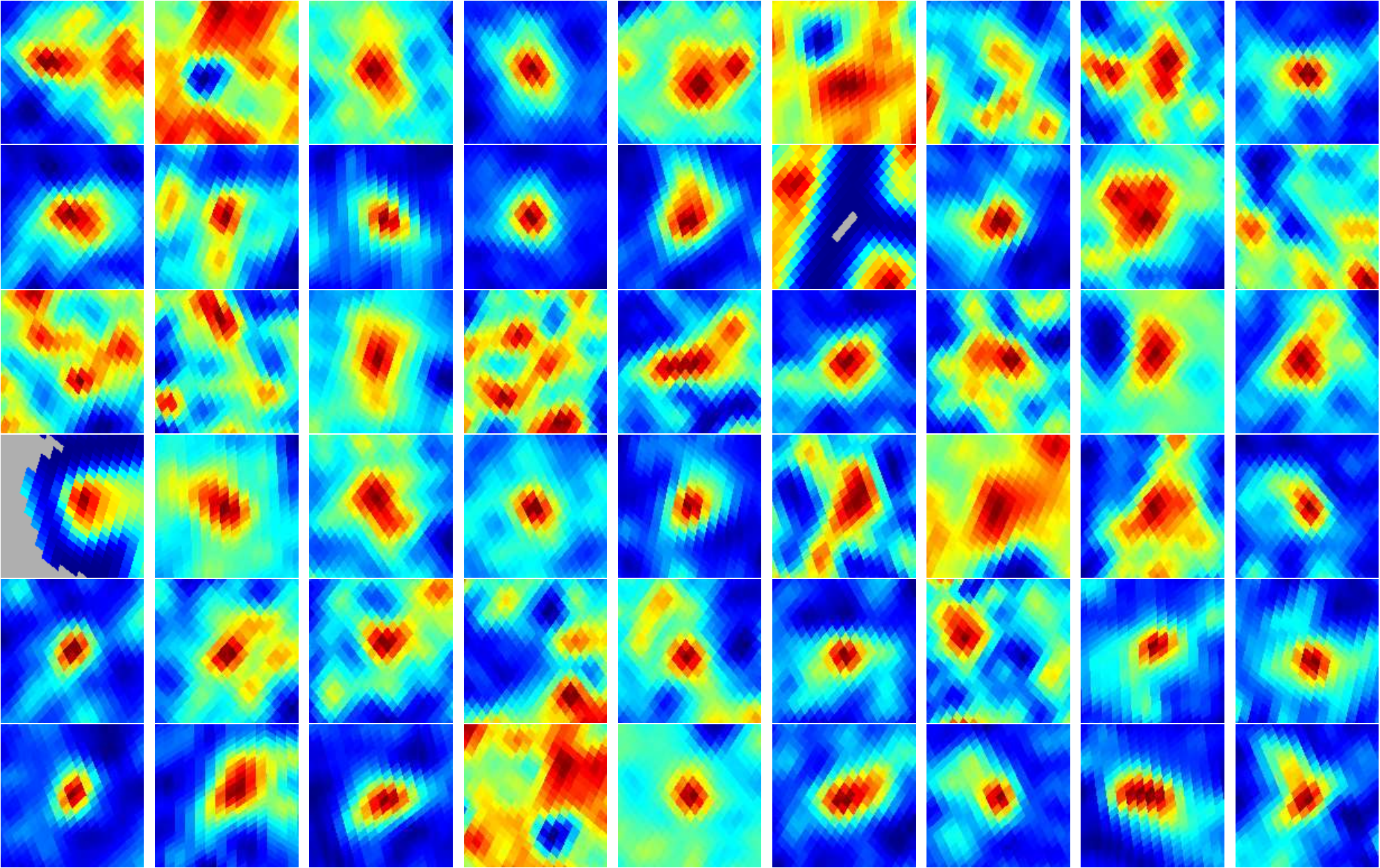}
\caption{Gnomic projection small maps of 54 individual clusters (sorted by decreasing signal-to-noise ratio given by \protect\cite{planckXXVII}) of the difference (NILC$-$2D-ILC) map (top 54 panels; \textit{upper half}) and the SDSS MphG map (bottom 54 panels; \textit{lower half}). Each map is a $0.5^{\circ}\times0.5^{\circ}$ field-of-view centred at the cluster position in the Galactic coordinate system with the north Galactic pole upwards. Grey areas in some of the stamps are masked regions of the sky.}
\label{fig:stamped}
\end{figure*}


\begin{figure}
\includegraphics[width=1.0\hsize]{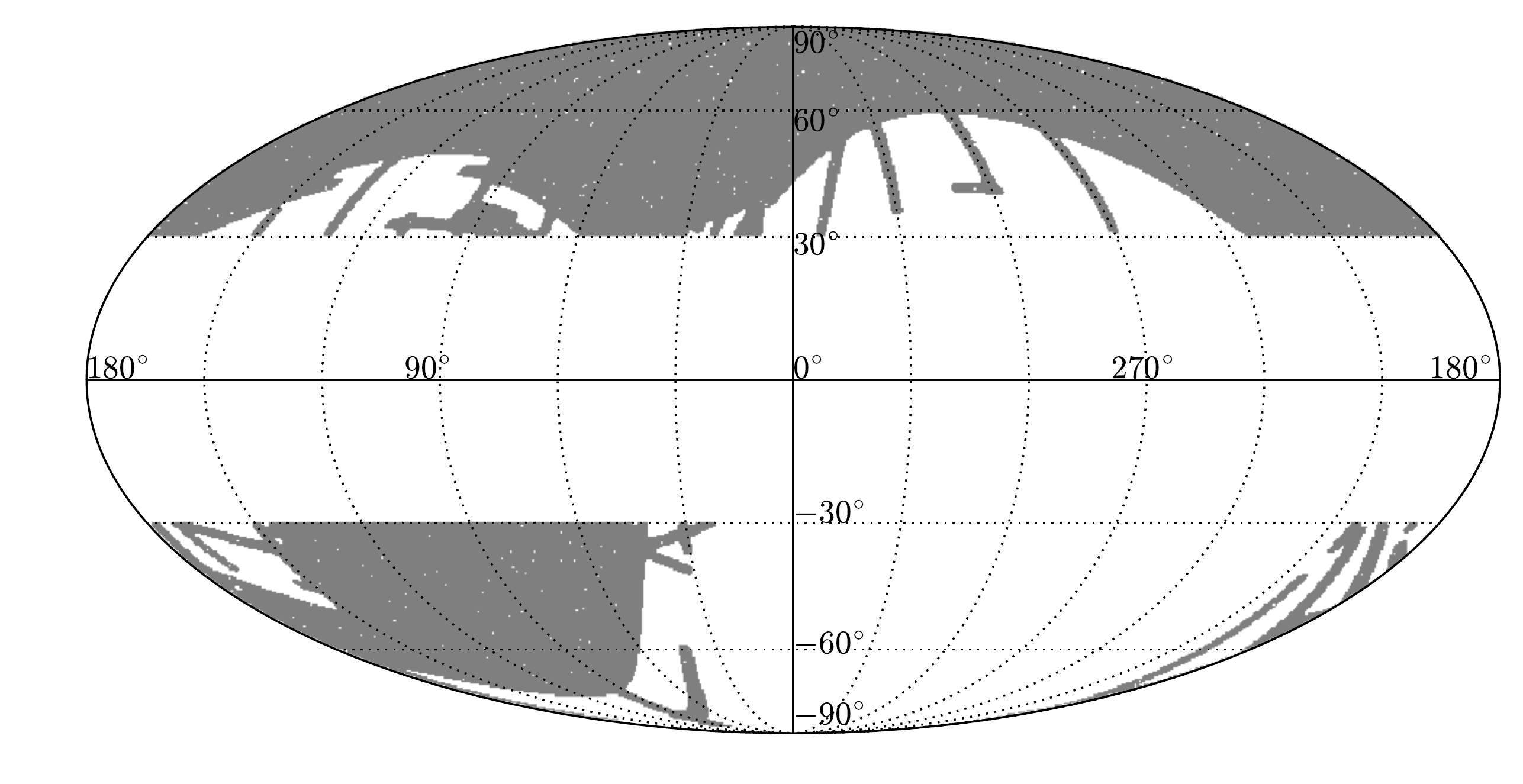}
\caption{The mask used in the power spectrum analysis.  The Galactic plane between the Galactic latitudes $|b|<30^{\circ}$ and the unobserved sky region of SDSS (\textit{white}) is masked out from our analysis.  The valid area (\textit{grey}) outside of the mask corresponds to $f_{\text{sky}}=0.27$.}
\label{fig:mask}
\end{figure} 

\subsection{Projection of individual clusters}
\label{stamper}

The spurious correlation between SZ cluster residuals in CMB maps and SDSS galaxies is also visible from the maps by looking at individual cluster locations in the sky. To highlight thermal SZ residuals in the \emph{Planck} NILC CMB map, we show in the top 54 panels of Fig.~\ref{fig:stamped} the difference (\mbox{NILC $-$ 2D-ILC}) map projected onto the locations of 54 selected clusters of the \emph{Planck} SZ catalogue \citep{planckXXVII} sorted by decreasing signal-to-noise ratio. The size of the stamps is of $0.5^{\circ}\times0.5^{\circ}$. Each stamp clearly shows negative (blue) temperature fluctuations at the position of the cluster due to residual thermal SZ emission in the \emph{Planck} NILC CMB map. Similarly, the bottom 54 stamps of Fig.~\ref{fig:stamped} show the SDSS MphG map smoothed to the same resolution of 5\,arcmin than the \emph{Planck} CMB map, and projected onto the same cluster locations in the sky. We see in this case an overdensity from SDSS galaxies at the positions of the SZ clusters of the \emph{Planck} SZ catalogue. These direct projections of the maps at the cluster positions gives a complementary view of the spurious anti-correlations between the SZ cluster residuals in CMB maps and galaxy densities in the SDSS survey. 

We recommend the use of the \mbox{2D-ILC} CMB map for CMB-LSS cross-correlation studies, given that this SZ-free CMB map appears to be safe from residual LSS foreground correlations. A more quantitative analysis on cross-power spectra is developed in the next section in order to corroborate our preliminary findings based on visual inspection of the maps.


\begin{figure}
\includegraphics[width=1.0\hsize]{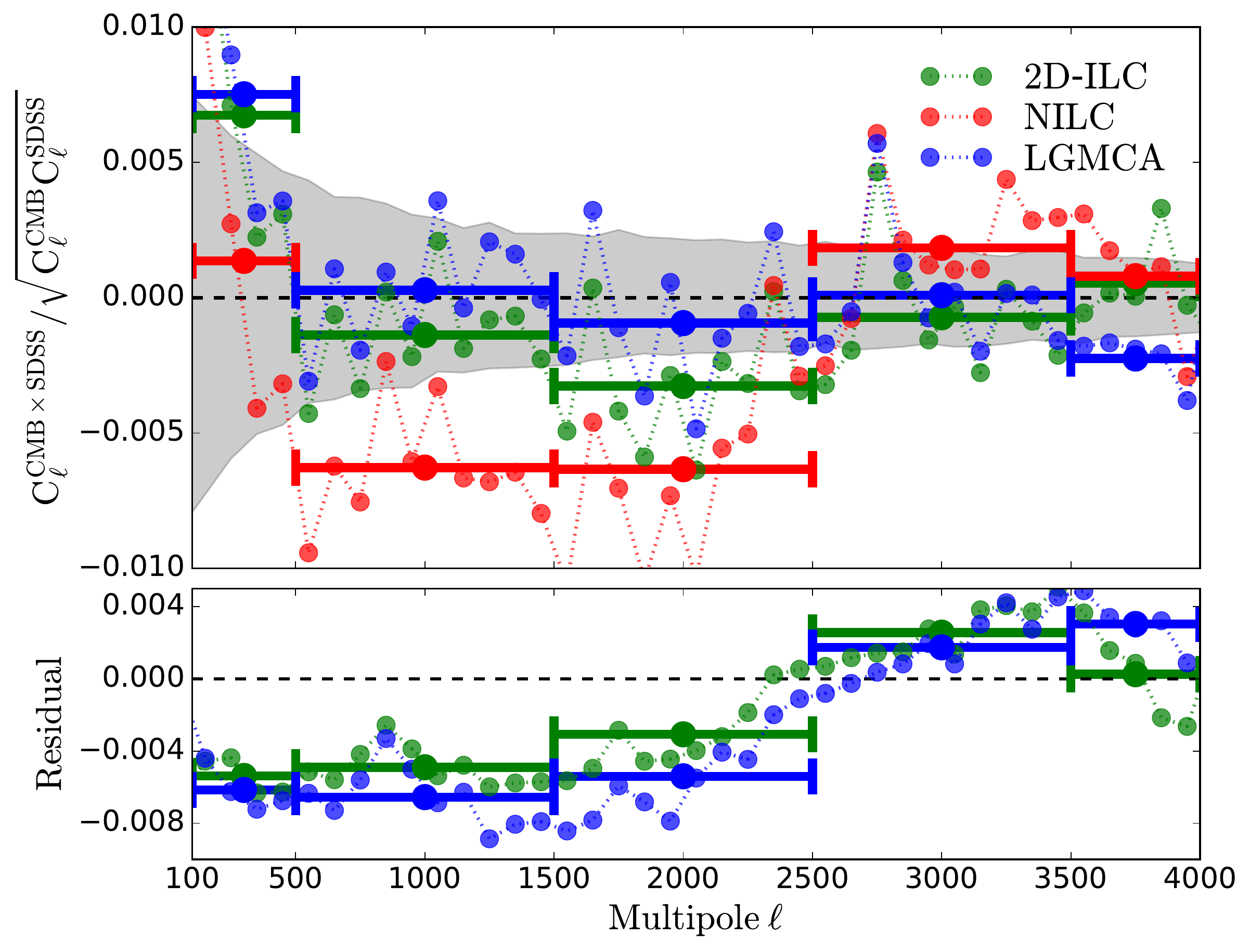}
\caption{\textit{Upper panel}: Cross-correlation coefficients of the SDSS MphG map with the \emph{Planck} 2015 NILC CMB map (\textit{red}),  the SZ-free 2D-ILC CMB map (\textit{green}) and the LGMCA map (\textit{blue}). The grey shaded area shows the $1\sigma$ uncertainty due to sample variance calculated from Monte Carlo simulations of pure CMB realisations. \textit{Lower panel}: the cross-correlation coefficient difference of the (SDSS $\times$ NILC) $-$ (SDSS $\times$ 2D-ILC) (\textit{green}) and (SDSS $\times$ NILC) $-$ (SDSS $\times$ LGMCA) (\textit{blue}).  Filled circles connected by dotted lines give the data binned within multipole ranges of $\Delta\ell = 100$, while thick horizontal bars give the data averaged over larger bin widths.}
\label{fig:cmbboss}
\end{figure} 

\section{Cross-power spectrum analysis} \label{cross_spec}

We use the \textsc{polspice} \citep{spc01,ccp+04,cc05} estimator to compute the angular cross-power spectra between the \emph{Planck} and SDSS survey maps. It does this by first computing the angular correlation function before transforming to the power spectrum. This allows additional apodization on large scales to reduce the aliasing of power from large to small scales in the presence of a mask.

To minimize the presence of residual contamination from Galactic foreground emission in the CMB maps, the Galactic plane is masked. We mask out all pixels at Galactic latitudes $|b|<30^{\circ}$, which mitigates the contamination from Galactic foregrounds while keeping enough signal for the cross-correlation analysis with SDSS. The unobserved sky region of SDSS is also masked out on the \textsc{healpix} map in addition to the CMB mask. Finally, the \emph{Planck} 143\,GHz intensity point source mask \citep{planckXI15} is used to avoid contamination from point sources.  The total masked region is shown as the white area in Fig.\,\ref{fig:mask}.  The mask is apodized by an 80\,arcmin beam to avoid spherical harmonic transform artefacts from the sharp boundary of the mask when computing angular power spectra. Afterwards, a threshold of 0.5 is imposed where pixels with a value below this threshold are set to zero. The sky coverage corresponds to 27\,\% ($f_{\text{sky}}=0.27$).

The maximum distance $\theta_{\mathrm{max}}$ used in the \textsc{polspice} estimator to integrate the correlation functions \citep[e.g.,][]{spc01} for power spectrum calculation is $40^\circ$. This value is chosen to minimise the effects of masking and because the available sky area means that the very largest angular scales ($\ell \lesssim 10$) are not reliably measured. We tested several values of $\theta_{\mathrm{max}}$, from $10^{\circ}$ to $80^{\circ}$ and found consistent results for $\ell > 100$. The scale factor of the correlation function tapering  \citep[e.g.,][]{spc01} is half of $\theta_{\mathrm{max}}$ (i.e., $20^\circ$).  The $f_{\text{sky}}=0.27$ factor due to the mask, the \textsc{healpix} $N_{\rm side}=2048$ pixel window function, and the 5\,arcmin beam convolution effect are all corrected for by \textsc{polspice}. The angular power spectrum values are computed into multipole bins of width varying from $\Delta\ell=100$ to $\Delta\ell=1000$, depending on the analysis.   

In order to quantify the (anti-)correlation between SZ cluster residuals in CMB maps and SDSS galaxies, we compute the dimensionless Pearson correlation coefficient over the angular scales as:
\begin{equation}\label{eq:ccc1}
c_\ell = {C_\ell^{\rm CMB \times SDSS} \over \sqrt{C_\ell^{\rm CMB}\, C_\ell^{\rm SDSS}}},
\end{equation}
where 'CMB' stands for either the \emph{Planck} 2015 NILC CMB map or the SZ-free 2D-ILC CMB map, and 'SDSS' stands for the SDSS MphG galaxy survey map smoothed to the same angular resolution of 5\,arcmin as the CMB maps.  
In order to focus on the thermal SZ residuals, we will also consider the correlation coefficient:
\begin{equation}\label{eq:ccc2}
\tilde{c}_\ell = {C_\ell^{\rm Diff \times SDSS} \over \sqrt{C_\ell^{\rm Diff}\, C_\ell^{\rm SDSS}}},
\end{equation}
where 'Diff' stands for the difference (NILC$-$2D-ILC) map. These dimensionless cross-spectra quantify the fractional correlated signal in the map.


\begin{figure}
\includegraphics[width=1.0\hsize]{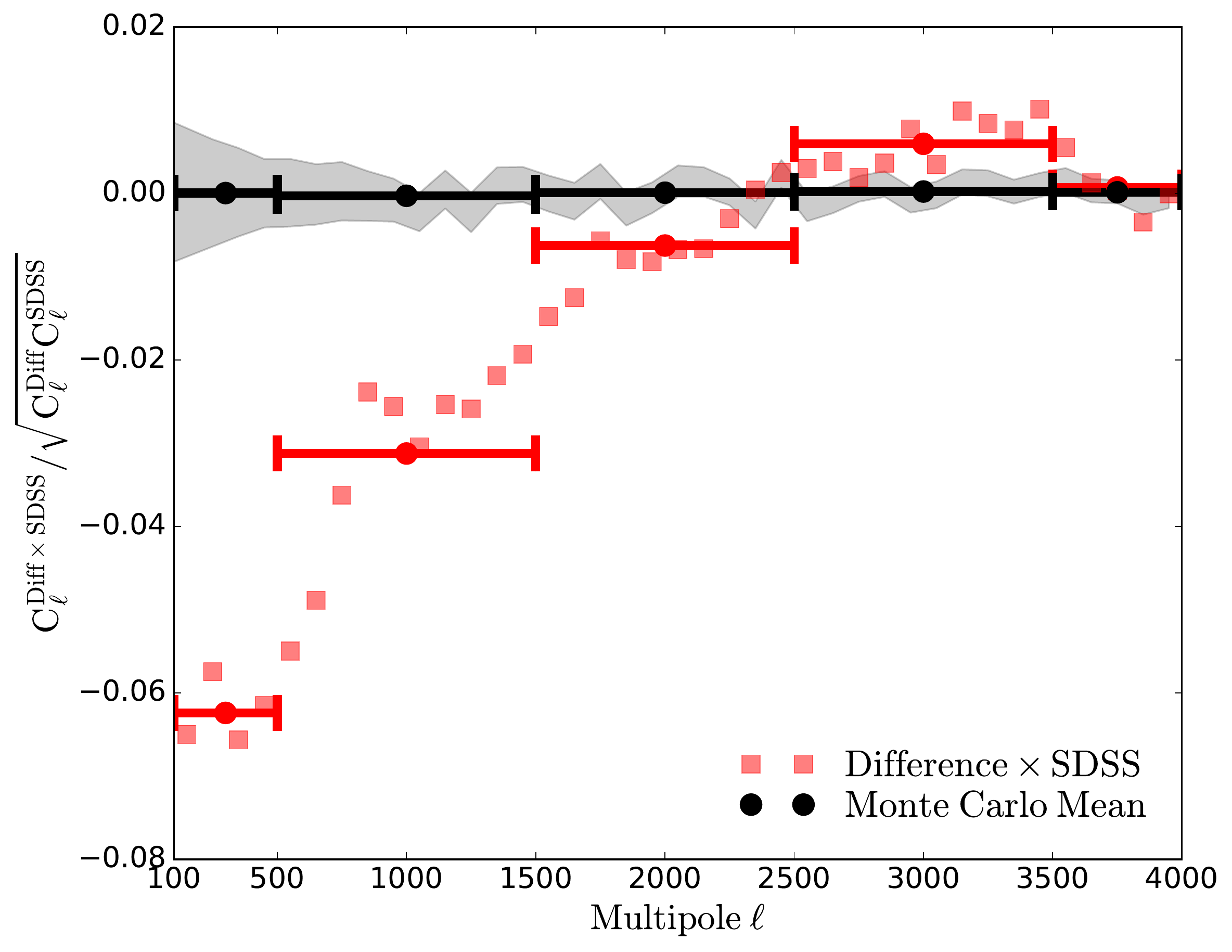}
\caption{Cross-correlation coefficients between the SDSS MphG map and the difference (NILC$-$2D-ILC) map (i.e., the SZ cluster residual map). The grey shaded area shows the $1\sigma$ uncertainty due to sample variance, calculated from Monte Carlo simulations with random positions of SDSS galaxies. Red squares give the data binned within multipole ranges of $\Delta\ell = 100$, while thick horizontal bars give the data averaged over larger bin widths.}
\label{fig:diffxboss}
\end{figure} 

\subsection{CMB $\times$ SDSS}
\label{sec-cmbxsdss}

We first compute the cross-correlation coefficient (Eq.~\ref{eq:ccc1}) between CMB and large-scale structures. Due to the ISW effect, the CMB temperature fluctuations and the distribution of large-scale structures are expected to show a positive correlation on the largest angular scales, corresponding to $\ell \lesssim 200$ (or $>1^{\circ}$) \citep[e.g.,][]{planckXXI}. However, CMB and large-scale structures must be theoretically uncorrelated on smaller angular scales $\ell > 500$, as long as the CMB map is not contaminated by large-scale structure foregrounds such as SZ clusters. 

In order to assess the level of sample uncertainty inherent to the cross-correlation coefficient statistics, we perform the following Monte Carlo (MC) simulation. We first simulate 1000 random Gaussian realisations of pure CMB maps from a theoretical CMB power spectrum generated by \textsc{camb} (\citealt{aa11}; \citealt{hlh+12}), based on the \emph{Planck} 2015 $\Lambda$CDM model \citep{planckXIII15}. We then compute the cross-correlation coefficient Eq.~\ref{eq:ccc1} between the 1000 random pure CMB realisations and the SDSS MphG map, so that the dispersion of the sample provides the $1\sigma$ uncertainty in each $\ell$ bin. 

The scale-dependent cross-correlation coefficient (Eq.~\ref{eq:ccc1}) computed between the SDSS MphG map and the NILC (resp. the 2D-ILC) CMB map is shown in red (resp. green) in the upper panel of Fig.~\ref{fig:cmbboss}. The $1\sigma$ sample variance based on the MC simulation is shown as the grey shaded region. The first bin of 100 multipoles ($\ell=2$--$100$) are removed from our analysis because the largest scale modes are significantly affected by the mask and apodization. The cross-correlation is thus divided into 5 bins of multipoles: $100<\ell<500$, $500<\ell<1500$, $1500<\ell<2500$,  $2500<\ell<3500$, and $3500<\ell<4000$. 

We detect a clear excess of anti-correlation at more than $1\sigma$ between the \emph{Planck} NILC CMB map and the SDSS galaxy survey (red) on the full range of angular scales between $\ell=500$ and $\ell=2500$. The observed anti-correlation between the \emph{Planck} NILC CMB map and SDSS galaxy density map is caused by the negative thermal SZ residual contamination from galaxy clusters in the CMB map that we have highlighted in Sect.~\ref{anasec}. In contrast, the 2D-ILC CMB map contains no thermal SZ effect by construction, and thus LSS-correlated residuals are minimised. The cross-correlation of the 2D-ILC CMB map with LSS from SDSS in Fig.\,\ref{fig:cmbboss} (green) shows less correlation on average, and is consistent with zero within $\sim 1\sigma$ sample variance, although for the multipole bin of $1500<\ell<2500$ the correlation is slightly larger than $\sim 1\sigma$.  This $1\sigma$ deviation from zero might be attributed to chance correlations with kinetic SZ residuals that do not average out due to sample selection, mis-calibration of thermal SZ (i.e. relativistic SZ effects which were neglected in the analysis), or other compact foreground residuals in the 2D-ILC map.  The level of anti-correlation between the \emph{Planck} NILC CMB map and the SDSS galaxies due to SZ residuals is not larger than $0.5\%$, nevertheless is significantly detected over a broad range of multipoles. Below $\ell=200$, the positive correlation between the \emph{Planck} NILC CMB map and SDSS is due to ISW effect but still might be underestimated because of large-scale SZ contamination.        

The lower panel of Fig.~\ref{fig:cmbboss} shows the difference between the NILC $\times$ SDSS and 2D-ILC $\times$ SDSS correlation coefficients (green), highlighting the scale-dependence of the anti-correlation between CMB SZ residuals and SDSS galaxies. The removal of the thermal SZ impacts a wide range of scales, from the largest angular scales down to $\ell \sim 2500$. For the first bin ($\ell=100$--500) both CMB maps do not show a significant correlation with LSS and therefore is probably dominated by Galactic and extragalactic foregrounds. In the range $\ell=500$--2500 the difference is attributed primarily due to the removal of the thermal SZ effect in the 2D-ILC map. Above $\ell \sim 2500$ there is no significant correlation due to the lack of angular resolution and the presence of noise in the CMB maps. However, we warn that for CMB maps measured at a higher resolution from, e.g., ACT \citep{faa+10} and SPT \citep{lrs+10}, one might see correlations at smaller scales.

The cross-correlation coefficient between the SDSS MphG map and the LGMCA CMB map is also computed for a consistency check as shown in blue in the upper panel of Fig.~\ref{fig:cmbboss}.  By comparing the blue and green lines, LGMCA gives consistent results with 2D-ILC such that the cross-correlation coefficients in both cases are consistent with zero within $\sim 1\sigma$ sample variance. As shown in blue in the lower panel of Fig.~\ref{fig:cmbboss}, similar to the 2D-ILC map, a clear excess of anti-correlation is present in the difference between the NILC $\times$ SDSS and LGMCA $\times$ SDSS correlation coefficients,  due to the thermal SZ residuals in the NILC CMB map. However, since the LGMCA CMB map combines both \emph{Planck} and \emph{WMAP} data, we use the 2D-ILC CMB map for the cross-spectrum calculation in the rest of this paper to be consistent with the released \emph{Planck} 2015 CMB maps.

 
 \begin{table}
\centering
\begin{tabular}{c|l}
\hline
\hline
Multipole Range & S/N  \\
\hline
$[100, 500]$  & $23.0$     \\
$[500, 1500]$  & $38.9$  \\
$[1500, 2500]$  & $30.4$   \\
\hline
$[100, 2500]$  & $54.5$   \\
\hline
\end{tabular}
\caption{Detection significance of the spurious correlation between SZ cluster residuals in the \emph{Planck} CMB map and SDSS galaxies over different ranges of angular scales. }
\label{szSN}
\end{table}
 
\subsection{SZ residuals $\times$ SDSS}
\label{sec-diffxboss}

In order to quantify the amount of spurious anti-correlation between SDSS galaxies and SZ cluster residuals in the \emph{Planck} NILC CMB map, we now consider the difference map between the NILC CMB map and the SZ-free 2D-ILC CMB map. The difference (NILC$-$2D-ILC) map is dominated by residual thermal SZ emission from galaxy clusters, while the CMB signal has been cancelled out by the difference. This difference map contains also residual foregrounds (Galactic and extragalactic) and noise due to the different processing of the \emph{Planck} data by the NILC and 2D-ILC algorithms, but at a negligible level compared to SZ residuals, at least outside of the mask (see e.g., middle panel of Fig.~\ref{fig:bossstack}). Following the same process described in Sect.~\ref{sec-cmbxsdss}, we now compute the cross-correlation coefficient (Eq.~\ref{eq:ccc2}) between the SDSS MphG map and the difference map over the angular scales. This can be considered a proxy for the cross-correlation coefficient between SDSS galaxies and SZ cluster residuals in the CMB. 

Figure~\ref{fig:diffxboss} shows the scale-dependent anti-correlation between the SDSS MphG map and the SZ difference map (red). In this case, the sample variance is computed as follows. We randomize the galaxy positions in the SDSS MphG catalogue and generate 1000 random LSS catalogues with randomly distributed galaxies. A simulated LSS map is created from each of the 1000 random catalogues. The 1000 simulated LSS maps are then cross-correlated with the difference map, giving the $1\sigma$ uncertainty due to sample variance that is shown as the grey shaded area in Fig.~\ref{fig:diffxboss}.

Figure~\ref{fig:diffxboss} confirms that the spatial (anti-)correlation with galaxy overdensities in SDSS comes from SZ cluster residuals in the \emph{Planck} CMB map. This spurious correlation is detected well beyond $1\sigma$ sample variance at the cluster scale, and with larger significance at large angular scales due to the clustering of SZ clusters. Again, the spurious correlation signal is weak in absolute value, with a few \% on cluster scales ($\ell=1500$--$2500$) and up to $\approx 6\%$ on large scales ($\ell=100$--$500$). Nevertheless, the correlation is detected at high significance across a wide range of scales. We consider this result as our best estimate of the amount of spurious correlation between SZ foregrounds in CMB maps and SDSS galaxies, given that any chance correlation due to CMB or Galactic foregrounds has been eliminated from the difference (NILC$-$2D-ILC) map. These results are consistent with those of Fig.~\ref{fig:cmbboss}, where the cross-correlation was computed between the CMB and SDSS maps. Interestingly, the spurious correlation due to SZ residuals is not dominant at the cluster scale but on large angular scales. We will interpret this overall trend in Sect.~\ref{sec:clustering}. 

We quantify the significance of the detection of the anti-correlation between CMB SZ cluster residuals and SDSS galaxies by computing the total signal-to-noise ratio (SNR) in each bin of multipoles as:
\begin{equation}
\mathrm{S\over N}\bigg\vert_{[\ell-\Delta_\ell/2,\ell+\Delta_\ell/2]} = \left[\sum_{\ell'=\ell-\Delta\ell/2}^{\ell+\Delta\ell/2}\left(\frac{c_{\ell'}}{\sigma_{\ell'}}\right)^2\right]^{1/2}\,,
\end{equation}
where $c_\ell$ is the value of the cross-correlation coefficient at multipole $\ell$ between SDSS and the SZ residual map, while $\sigma_\ell$ is the corresponding $1\sigma$ sample variance from the Monte Carlo simulations at the same multipole. The SNR results for different ranges of angular scales are listed in Table\,\ref{szSN}. At the cluster scale ($1500 < \ell < 2500$), the anti-correlation is detected with $\approx 30\sigma$ significance, while overall the detection significance is $\approx 54\sigma$ over the full range of angular scales from $\ell=100$ to $\ell=2500$.  However, it should be mentioned that in addition to the contribution from SZ cluster residuals, the correlated signal might also hide a small contribution from other foreground residuals in the difference map due to inherent differences between the two processed CMB maps. Indeed, when we compute the cross-correlation coefficient between the difference map and the GNILC {\it Planck} thermal dust map at 545\,GHz and 857\,GHz \citep{planckXLVIII}, we observe a positive correlation below $\ell = 300$ and a negative correlation at $300<\ell<3000$. This negative correlation on cluster scales will contribute to the cross-correlation coefficient in Fig.\,\ref{fig:diffxboss}, resulting in a stronger detection SNR. The positive correlation observed between the difference map and foreground dust maps below $\ell = 300$ might explain the positive bias in the first $\ell$-bin ($\ell=100$--500) of Fig.\,\ref{fig:cmbboss}, although the positive bias is at the $\sim 1\sigma$ level.


\begin{figure}
\includegraphics[width=1.0\hsize]{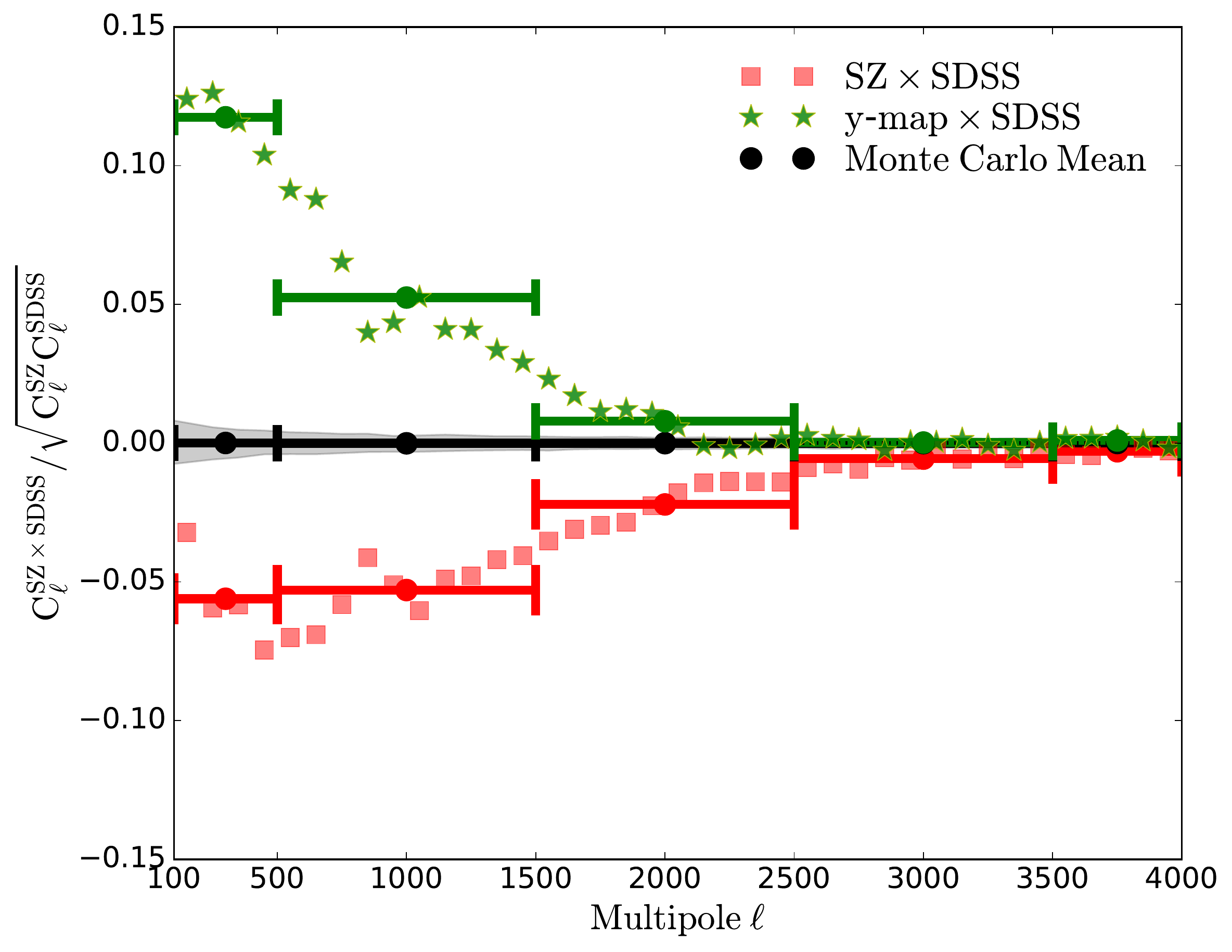}
\caption{Cross-correlation coefficient between the SDSS MphG map and (i) the catalogue SZ cluster map at $143$\,GHz created from the \emph{Planck} 2015 SZ catalogue (\textit{red}; \protect\citealt{planckXXVII}), (ii) the  \emph{Planck} NILC thermal SZ $y$-map (\textit{green}, \citealt{planck2015sz}).  The \emph{Planck} SZ $y$-map has a beam resolution of 10\,arcmin, so that the SDSS MphG map has been smoothed accordingly down to 10\,arcmin in that case for the cross-correlation.  The $1\sigma$ smaple variance calculated from the Monte Carlo simulations is shown as the grey shaded area. Discrete points give the averaged data over multipole bins of width $\Delta\ell = 100$, while thick horizontal bars give the data averaged over larger bin widths. }
\label{fig:szxboss}
\end{figure} 

\begin{figure}
\includegraphics[width=1.0\hsize]{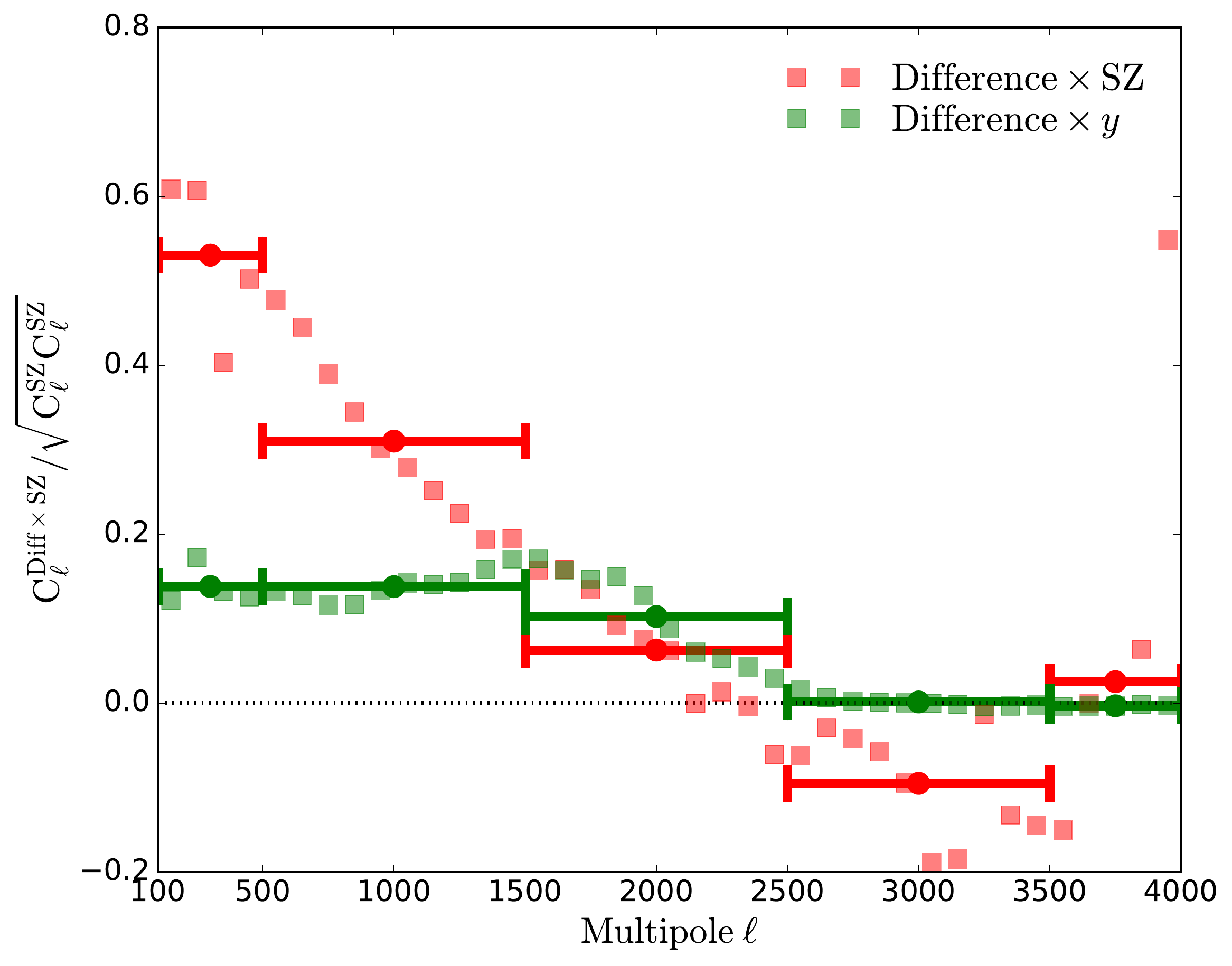}
\caption{Cross-correlation coefficient between the difference map and (i) the catalogue SZ cluster map at $143$\,GHz created from the \emph{Planck} 2015 SZ catalogue (\textit{red}; \protect\citealt{planckXXVII}), (ii) the  \emph{Planck} NILC thermal SZ $y$-map converted into temperature at 143 GHz (\textit{green}, \citealt{planck2015sz}).  The \emph{Planck} SZ $y$-map has a beam resolution of 10\,arcmin, so that the difference map has been smoothed accordingly down to 10\,arcmin in that case for the cross-correlation. Discrete points give the averaged data over multipole bins of width $\Delta\ell = 100$, while thick horizontal bars give the data averaged over larger bin widths. }
\label{fig:leftover}
\end{figure}

\subsection{SZ $\times$ SDSS}
\label{sec-szxboss}

To investigate further the correlation of the residuals with SDSS,  it is instructive to generate a pure SZ cluster map. We use the catalogue SZ map at $143$\,GHz (which is the frequency at which the bulk of SZ cluster residuals in the CMB maps comes from) from the \emph{Planck} SZ catalogue, as described in Sect.~\ref{szcat}.  The cross-correlation between the SDSS MphG map and the catalogue SZ map at 143\,GHz is shown in red in Fig.~\ref{fig:szxboss}. The grey area shows the $1\sigma$ sample variance, computed by generating 1000 SZ maps obtained by randomizing the locations of the SZ clusters in the \emph{Planck} 2015 SZ catalogue, while keeping the Compton parameter fluxes from the catalogue. The catalogue SZ map at $143$\,GHz confirms the overall trend of anti-correlation with SDSS galaxies, as observed from the data. Of course, in this case the correlation coefficient is larger because the full thermal SZ signal contributes to the correlation signal, while on the data only a fraction of the thermal SZ emission is present as a residual contamination in the \emph{Planck} CMB maps. 

As further evidence, we also cross-correlate the \emph{Planck} 2015 NILC thermal SZ y-map \citep{planck2015sz}, based on \emph{Planck} sky observations, with the SDSS MphG map (green curve in Fig.~\ref{fig:szxboss}). The \emph{Planck} SZ $y$-map has a \textsc{healpix} pixel resolution of $N_{\rm side} = 2048$ but a beam resolution of 10\,arcmin. In order to cross-correlate the \emph{Planck} $y$-map with the SDSS MphG map, we thus smooth the SDSS MphG map down to 10\,arcmin. Since the $y$-Compton parameter is positive, the cross-correlation coefficient of the \emph{Planck} SZ $y$-map with the galaxy overdensities of the SDSS MphG map is also positive. In any case, we recover the general trend observed in the cross-correlations between the \emph{Planck} CMB map and the SDSS survey map, which confirms that the observed spurious anti-correlation is due to SZ cluster residuals in the \emph{Planck} CMB maps.

Both cross-correlation coefficients in Fig.~\ref{fig:szxboss} show that there is a correlation between SZ clusters and SDSS galaxies both in the typical cluster scale at $\ell\sim2000$, as expected from the agglomeration of galaxies within clusters, but mostly on large angular scales $\ell < 1500$, that we attribute to the clustering of SZ clusters. 

In order to estimate the amount of thermal SZ emission that has been left over in the \emph{Planck} 2015 NILC CMB map, we calculate the cross-correlatation coefficient of 
\begin{equation}
\delta = {C_\ell^{\rm Diff \times SZ} \over \sqrt{C_\ell^{\rm sz}\, C_\ell^{\rm sz}}},
\end{equation}
which gives an estimated percentage of the thermal SZ emission that has been left over in the \emph{Planck} 2015 NILC CMB map. We calculate $\delta$ using the 143 GHz catalogue SZ map and the \emph{Planck} $y$-map converted into temperature at 143 GHz respectively. The results are shown in Fig.\,\ref{fig:leftover}. At the typical cluster scale of $\ell\sim2000$, both maps give an estimated $10\%$ left-over of the thermal SZ emission in the \emph{Planck} NILC CMB map. On large scales, the estimated left-over based on the catalogue SZ map ($\sim 40\%$) is higher than that based on the \emph{Planck} $y$-map ($\sim15\%$). The difference between the two maps might be due to the foreground contamination in the \emph{Planck} $y$-map, which smears out some of the SZ signal and thus results in a smaller percentage when estimating the left-over of the thermal SZ emission in the \emph{Planck} NILC CMB map. 

\begin{figure}
\includegraphics[width=1.0\hsize]{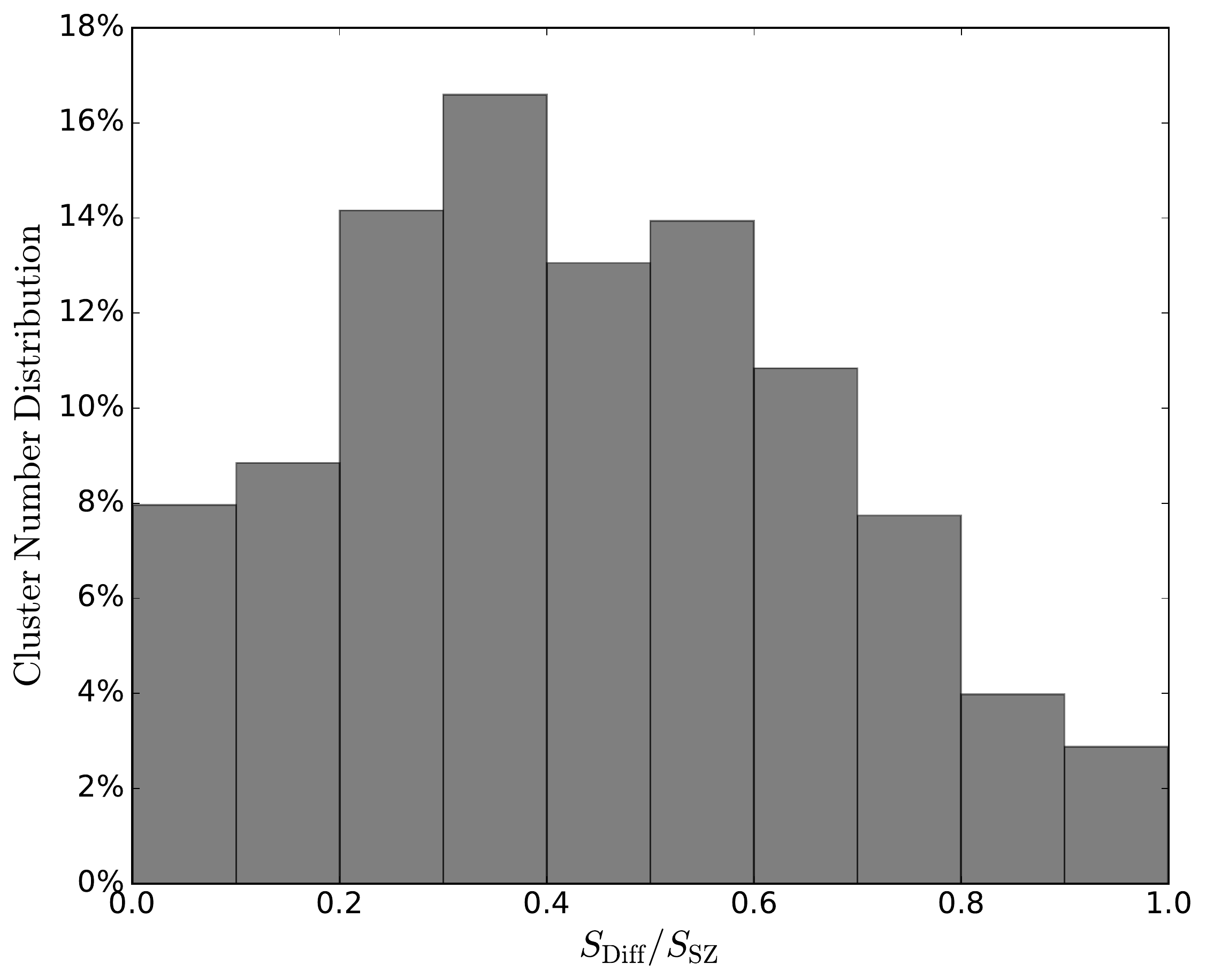}
\caption{The cluster number distribution as a function of the left-over percentage of the thermal SZ emission in the \emph{Planck} 2015 NILC CMB map. The $x$-axis gives the ratio of the integrated flux density of the SZ clusters from the difference map over that from the catalogue SZ cluster map, calculated using aperture photometry. The $y$-axis gives the number of sources in the form of the percentage over total number (452) of sources included in our analysis. The mean value is 40\%. }
\label{fig:histplot}
\end{figure}

The percentage of the left-over thermal SZ emission in the \emph{Planck} NILC CMB map is further quantified through aperture photometry. At each SZ cluster location given by the \emph{Planck} SZ catalogue, a circle with a radius of 1.5 times the cluster radius convolved  with the 5\,arcmin beam is used to calculate the integrated flux density within the circle.  Two annuli with respective radii of 1.8 and 2.2 times the cluster radius convolved with the beam are used to subtract the background flux density from the foreground circle, giving the integrated flux density of the cluster in the foreground circle. The scales of the foreground and background radii are chosen such that the foreground circle fully covers the cluster signal, taking into account the smearing effect from the beam at the edge of the clusters, and the background annuli is  large enough to not contain any foreground signal but small enough to contain only the local background noise.  The total number of SZ clusters from the \emph{Planck} SZ catalogue outside of the mask region included in our analysis is 549. The integrated flux density given by aperture photometry   of the included SZ clusters are calculated from both the difference map and the catalogue SZ cluster map, the ratio of which gives the percentage of the SZ emission that has been left in the \emph{Planck} NILC CMB map. Among the 549 included SZ clusters, 97 ($18\%$) give a ratio of either larger than 1 or smaller than 0, due to the noisy background in the difference map which affects the performance of aperture photometry.  We further discard these 97 noisy clusters from our statistics and the left-over percentage distribution of the rest 452 ($82\%$) clusters are plotted in Fig.~\ref{fig:histplot}. Over half ($58\%$) of the clusters give a ratio between 0.2 and 0.6 and overall give an average of 0.44, meaning that $44\%$ of the thermal SZ emission has been left over in the \emph{Planck} NILC CMB map. To test the robustness of this result we also split the 452 clusters into 5 groups of decreasing detection S/N ratio and found mean values of  0.488, 0.465, 0.413, 0.394 and 0.410, respectively. This shows that the mean residual flux is robust with a standard deviation of 0.04. This is consistent with the cross-spectrum analysis using the difference map and the catalogue SZ cluster map (red) in Fig.~\ref{fig:leftover} which suggests $\sim 30$--60\% on scales $\ell=100$--1000. Therefore, we consider $44\pm4\%$ as our best estimate for the percentage of thermal SZ emission that has been left over in the \emph{Planck} NILC CMB map.


\begin{figure}
\includegraphics[width= \hsize]{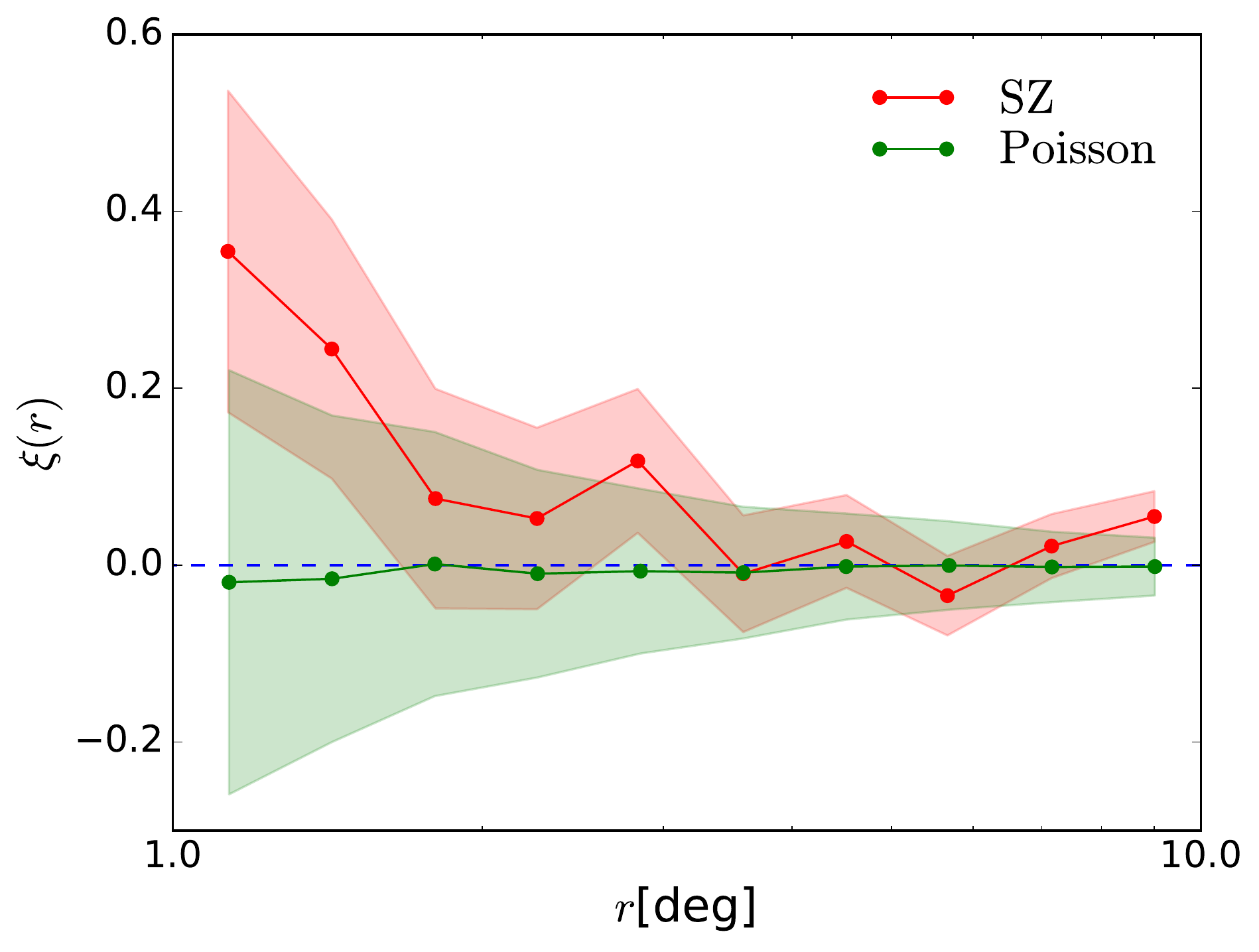}
\caption{Two-point angular correlation function of the \emph{Planck} 2015 SZ catalogue (\textit{red}) and a random catalogue with Poisson distributed sources (\textit{green}). The $1\sigma$ sample variance calculated from Monte Carlo simulations in each case is shown as the shaded areas.}
\label{fig:2pointcorr}
\end{figure}

\subsection{Large-scale anti-correlation from clustering of SZ cluster residuals}\label{sec:clustering}

Figures~\ref{fig:cmbboss}, \ref{fig:diffxboss}, and \ref{fig:szxboss} all show large anti-correlation signals at large angular scales, $100 < \ell < 1500$, which disappears in the Monte Carlo simulations where source locations have been randomized. Therefore, the large-scale anti-correlation signal cannot be attributed to artefacts due to masking and apodization. In order to understand large-scale anti-correlation, we compute the two-point correlation function of the thermal SZ emission as a function of angular separation over the sky to confirm the detection of large-scale clustering of SZ clusters. We use the \textsc{TreeCorr} package (\citealt{jbj04}) to compute the two-point correlation function as
\begin{equation}\label{eq:xi}
\xi = \frac{DD-2DR+RR}{RR}\,,
\end{equation}
where $DD$ is the counts of pairs of SZ clusters in the \emph{Planck} 2015 SZ catalogue as a function of separation $r$ for each bin, $RR$ is the counts of pairs in a random catalogue with Poisson distributed sources and $DR$ is that between the SZ catalogue and the random catalogue (\citealt{ls93}).

The minimum and maximum separation is chosen to be 1\,degree and 10\,degrees respectively, split into 10 bins on a logarithmic scale.  We create 1000 random catalogues of Poisson distributed sources and compute the two-point correlation (Eq.~\ref{eq:xi}) for the SZ catalogue using each random Poisson catalogue, the mean of which is plotted as the red curve in  Fig.~\ref{fig:2pointcorr}, and the dispersion of which provides the $1\sigma$ uncertainty shown as the red-shaded area.  For comparison, we also compute the two-point correlation among the 1000 random Poisson catalogues shown as the green curve in Fig.~\ref{fig:2pointcorr}. In each case, the larger uncertainty towards smaller separations is because, for each source, one annulus with a width equal to the bin width is used to compute the correlation function for that bin. Therefore, a larger separation corresponds to a larger area with more samples for calculating the correlation function and thus gives less sample variance.  While the two-point correlation function for the random Poisson catalogues is in average zero over the full range of angular separations over the sky, the two-point correlation function for the SZ catalogue shows positive peaks of correlation on degree scales, thus detecting the effect of large-scale clustering of SZ clusters at $3.3\sigma$. 

This result confirms that the anti-correlation signal at scales beyond the typical cluster ($\ell \approx 100$--1000) is mainly caused by the clustering of SZ galaxy cluster residuals in CMB maps on scales of a few degrees.


\begin{figure}
\includegraphics[width=1\hsize]{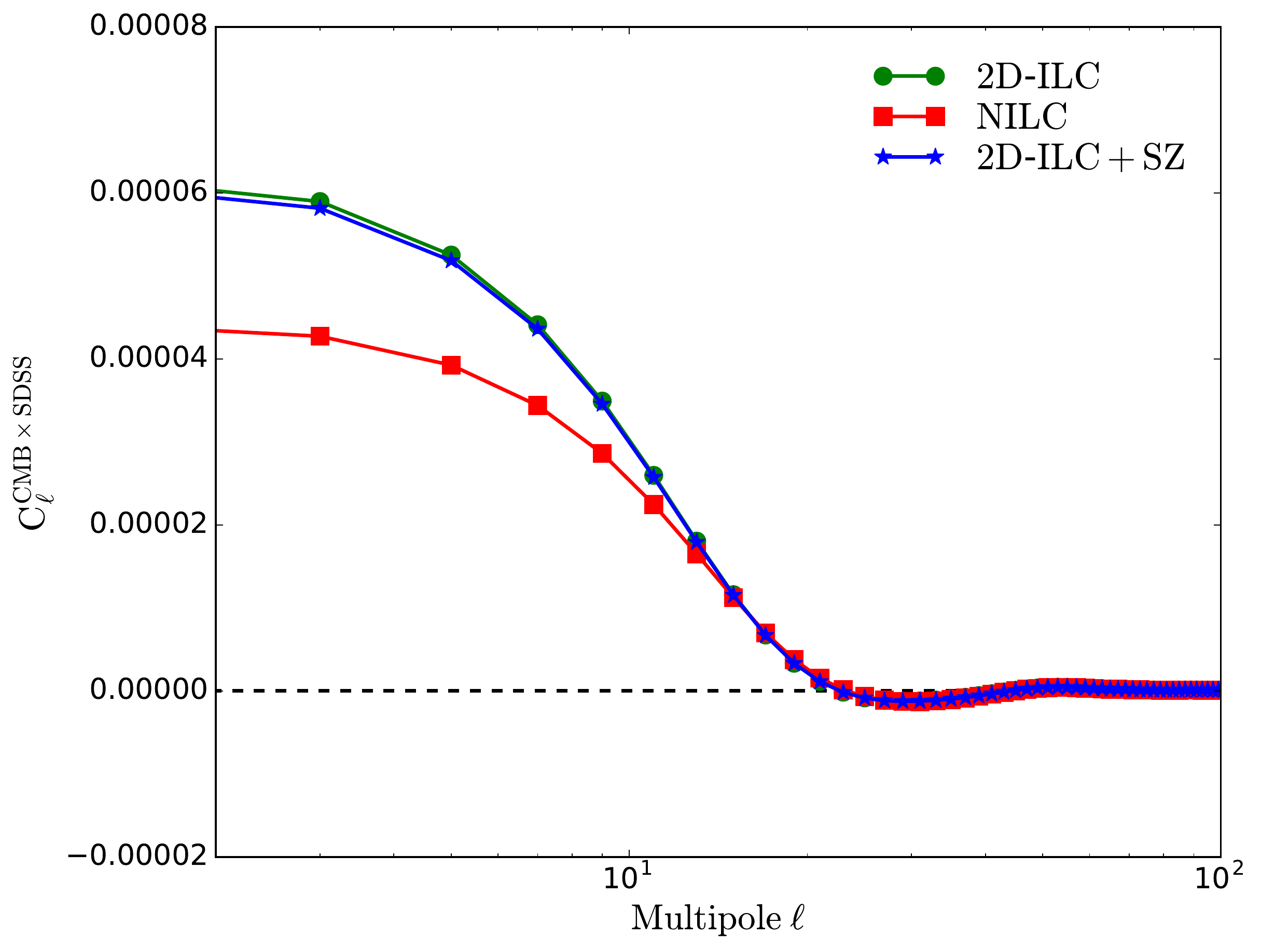}
\caption{Cross-power spectra between the SDSS MphG density contrast map and the  NILC CMB map (\textit{red}), the 2D-ILC CMB map (\textit{green}), and an artificial SZ-contaminated CMB map (\textit{blue}), where the catalogue SZ 143\,GHz map has been added in to the SZ-free 2D-ILC CMB map.}
\label{fig:szisw}
\end{figure} 

\subsection{Impact on the measurement of the Integrated Sachs-Wolfe effect}
\label{sec-isw}

Given that SZ cluster residuals in the \emph{Planck} CMB map create a spurious anti-correlation signal in the cross-power spectrum between CMB and SDSS maps, this could potentially be an issue for the measurement of the ISW effect by cross-correlation. Indeed the amplitude of the positive correlation signal due to ISW effect might be underestimated because of the competing anti-correlation signal due to SZ residuals in the CMB map. 

In \cite{planckXIX} and \cite{planckXXI}, the cross-power spectrum between the \emph{Planck} CMB maps and the SDSS MphG density contrast map was computed to measure the ISW effect at large angular scales. In order to quantify the impact of the SZ cluster residuals on the ISW detection under the same conditions as in \cite{planckXXI}, in Fig.~\ref{fig:szisw} we instead compute the cross-power spectrum between the SDSS MphG map and (i) the \emph{Planck} NILC CMB map (\textit{red}), (ii) the SZ-free 2D-ILC CMB map (\textit{green}) on large angular scales. We see that the spectra are almost identical, with up to a $\approx40\%$ difference relative to the NILC CMB map in the power on the largest angular scales ($\ell < 10$).

To understand whether the gap between the two ISW measurements comes either from SZ cluster residuals in the \emph{Planck} NILC CMB map or from stronger residual contamination from Galactic foregrounds in the SZ-free 2D-ILC CMB map, we artificially include SZ contamination into the 2D-ILC CMB map, by adding the catalogue SZ 143\,GHz map to the SZ-free 2D-ILC CMB map. We calculate the cross-power spectrum between this artificial map and the SDSS MphG  map as shown in Fig.~\ref{fig:szisw} (\textit{blue}). Compared with the result from the SZ-free 2D-ILC CMB map (green), the artificial SZ contamination indeed reduces the amplitude of the cross-power spectrum by $\approx1.7\%$ due to the anti-correlation between the SZ cluster contamination and SDSS galaxies.  However, the gap in amplitude between the SZ-free 2D-ILC (green) and NILC (red) maps is a factor of $\approx 17$ larger than  this artificially-added anti-correlation. Given that the artificial SZ contamination in this case is the full power of the SZ emission from the \emph{Planck} 2015 SZ catalogue while the actual SZ cluster contamination in the \emph{Planck} NILC CMB map is $\sim 40\%$, the impact on the ISW measurement appears to be insignificant. 

This result is somewhat expected from the absolute value of the correlation coefficients between SDSS and CMB SZ residuals that we have measured in Fig.~\ref{fig:diffxboss} to be less than 6\% at large angular scales. This is consistent with \cite{als04}, where by cross-correlating WMAP with 2MASS, they found that the ISW and SZ components dominate before and after $\ell = 20$ respectively. The removal of SZ cluster residuals in the 2D-ILC CMB map is at the cost of more contamination from Galactic foregrounds. Therefore, the excess of amplitude in the CMB-SDSS cross-power spectrum at large angular scales observed for the 2D-ILC map might result from chance correlation with stronger residual Galactic foreground contamination. Nevertheless, \cite{mh18} shows that using 2D-ILC on the CMB lensing potential map will allow for robust SZ-free CMB lensing measurement without much penalty caused by the increased global noise.


\subsection{Impact on CMB lensing-galaxy lensing correlations}

Residual thermal SZ contamination of CMB maps by galaxy clusters will propagate to the CMB lensing potential map, given that it is derived from the contaminated CMB map itself by mean of quadratic estimators. Lensing quadratic estimators \citep[e.g.,][]{Hu2002} perform a weighted convolution of CMB maps in spherical harmonic space in order to reconstruct the lensing potential $\Phi$:
\begin{equation}
\widehat{\Phi}(\Lb) \propto \int {d^2\ell'\over (2\pi)^2}\, W(\Lb,\ellb)\, T^{\rm CMB}(\ellb)\,T^{\rm CMB}(\Lb - \ellb),
\end{equation}
where $T^{\rm CMB}(\ellb)$ is the spherical harmonic coefficient of the CMB map at multipole $\ellb$, and
\begin{equation}
W(\Lb,\ellb) = {\Lb \cdot \ellb C_\ell + \Lb \cdot \left(\Lb-\ellb\right) C_{\vert \Lb-\ellb \vert} \over \left(C_\ell + N_\ell \right)\left(C_{\vert \Lb-\ellb \vert}+N_{\vert \Lb-\ellb \vert}\right)},
\end{equation}
with $C_\ell$ and $N_\ell$ being the CMB and noise angular power spectra.

As discussed in this work, the \emph{Planck} CMB maps contains SZ cluster residuals, which in the case of the \emph{Planck} NILC CMB map are given by Eq.~\ref{szres}. Therefore, SZ cluster residuals will propagate to the CMB lensing potential as an additive bias:
\begin{equation}\label{reslens}
\Delta\widehat{\Phi}(\Lb)  \propto \left(\sum_\nu w_\nu b_\nu\right)^2\int {d^2\ell'\over (2\pi)^2}\, W(\Lb,\ellb)\, y(\ellb)\,y(\Lb - \ellb),
\end{equation}
where $w_\nu$ are the NILC weights across frequencies, $b_\nu$ is the SED of thermal SZ effect, and $y(\ellb)$ is the spherical harmonic coefficient of the $y$-Compton parameter of the thermal SZ effect at multipole $\ellb$. Equation~\ref{reslens} shows that lensing quadratic estimators mix SZ cluster residuals from different scales because of the convolution in harmonic space, so that the impact of SZ cluster residuals on the CMB lensing potential is not trivial, but might be more significant than for the ISW effect.

When performing CMB lensing-galaxy lensing cross-correlations \citep[e.g.,][]{als+15}, the SZ residuals in the CMB lensing potential will thus correlate with the lens sources of optical galaxy surveys through a bispectrum of the form:
\begin{equation}\label{bispec}
\langle y(\ellb),\, y(\Lb - \ellb),\, T^{\rm galaxy\, lenses} (\Lb)\rangle,
\end{equation}
which does not vanish because of the non-Gaussian distribution of large-scale structures in the sky and may lead to a very large bias \citep{van_engelen2014}. 
The detailed estimation of the spurious correlations in CMB lensing-galaxy lensing cross-correlations due to SZ cluster residuals in the CMB lensing map is beyond the scope of this paper. Nevertheless, the formal evidence for SZ cluster residuals in the CMB lensing potential map deserves a detailed investigation of the data. While we were completing this work, \cite{mh18} submitted a paper discussing similar issues of SZ residuals in CMB lensing maps, and the necessity to project SZ cluster residuals out of CMB maps. In particular, the authors have found that the bispectrum in Eq.~\ref{bispec} due to SZ residuals correlating with galaxies contributes a significant bias on the halo mass measurement. This shows the importance of ensuring that correlated residual contamination in maps are minimised as much as possible.


\section{Conclusions}\label{dissec}

We have quantified the amount of residual thermal SZ contamination from galaxy clusters in the \emph{Planck} 2015 CMB maps through stacking analysis and cross-correlation with the SDSS survey of galaxies. By cross-correlating the \emph{Planck} NILC CMB map with the SDSS MphG map, we have detected spurious anti-correlation with high significance over a large range of angular scales due to SZ cluster residuals in the \emph{Planck} CMB maps. At the cluster scale ($\ell\sim2000$), anti-correlation between SZ clusters residuals in CMB and SDSS galaxies is detected with $\approx 30\sigma$ significance, while by including large angular scales, we obtain an overall $\approx 54\sigma$ detection of the spurious anti-correlation signal. 

Residual thermal SZ emission at large angular scales in the CMB has already been discussed in the literature. For example, \cite{als04} had to fit for thermal SZ contamination in the \emph{WMAP} CMB map over a large range of scales from $\ell > 20$ when cross-correlating the \emph{WMAP} CMB map with the 2MASS survey in order to measure the ISW effect. These evidences further validate that the large-scale anti-correlation observed in Figs.~\ref{fig:cmbboss}, \ref{fig:diffxboss}, and \ref{fig:szxboss} is caused by the residual clustering of SZ clusters in the \emph{Planck} CMB maps. 

By comparison, the 2D-ILC CMB map that we have produced by using the Constrained ILC component separation technique shows negligible SZ cluster residuals. The high level of detection from large angular scales has to be put in perspective, given that part of the anti-correlation signal can be due to other residual foregrounds when taking the difference between the \emph{Planck} CMB map and the SZ-free 2D ILC CMB map. While detected with high significance, the spurious correlation between CMB SZ cluster residuals and SDSS galaxies is not larger than 6\%. Note that we neglect relativistic corrections to the SZ signal \citep[e.g.,][]{cl98, ikn98, nik98, nis+06, cns+12,  hgl+18}. The ensemble averaged SZ temperature of {\it Planck} clusters is found to be $kT_{\rm e}\simeq 4$--$6\,{\rm keV}$ \citep{ebc+18}, which means that relativistic temperature corrections are $\lesssim 5-10\%$, causing only a small possible contamination due to our treatment. Although kSZ component is left in the CMB due to spectral degeneracy between CMB and kSZ,  kSZ consists of both positive and negative fluctuations depending on the peculiar velocities of the clusters, which averages out to zero in an homogeneous universe. Therefore, we also neglect the correlations due to kSZ component as the cross-power spectrum between kSZ fluctuations and LSS will vanish out.

We estimate the percentage of the thermal SZ emission that has been left over in the \emph{Planck} 2015 NILC CMB map by (i) calculating the ratio of the integrated flux density at the SZ cluster locations between the difference map and an SZ cluster map created from the \emph{Planck} 2015 SZ catalogue \citep{planckXXVII} using aperture photometry, and (ii) by calculating the cross-correlation coefficient between the difference map and the catalogue SZ cluster map. Both methods give consistent results with our best value coming from the aperture photometry analysis of $44\pm4\%$ of the thermal SZ contamination remaining in the \emph{Planck} 2015 NILC CMB map.

The impact on measurements of the ISW effect of SZ cluster residuals in CMB maps at large angular scales has been calculated to be negligible, which puts confidence in the ISW detection made by \cite{planckXXI} by cross-correlation. However, the detected SZ cluster residuals in the \emph{Planck} CMB maps on other angular scales, and the spurious anti-correlation with LSS survey maps that they cause, can no longer be ignored in any cross-correlation analysis between CMB and LSS surveys. In particular, we warn that residual SZ clusters in CMB (and CMB lensing) maps, which correlate with the galaxy density (and shear) fields of optical surveys, should systematically be taken into account in any statistical interpretation of CMB-LSS cross-correlation measurements. In this context, we also recommend to use SZ-free CMB maps, such as the 2D-ILC map built in this work, for deriving the CMB lensing potential map and for cross-correlations with LSS surveys.

As pointed out in \cite{mh18}, SZ cluster residuals in CMB maps will also affect the CMB lensing potential field reconstruction by propagation through lensing quadratic estimators. In this case, lensing quadratic estimators should be better applied to an SZ-free CMB map, such as the 2D-ILC map, in order to avoid for example spurious correlations between the CMB lensing potential map and LSS optical survey maps. 

Finally, we warn that the SZ cluster residuals might be more of a problem for CMB maps observed with a higher resolution instrument, such as ACT \citep{faa+10}, SPT \citep{lrs+10} and the upcoming CMB-S4 experiment \citep{CMBS4}. In these cases, the anti-correlation caused by the thermal SZ cluster residuals when cross-correlating with LSS will have an effect on even smaller scales ($\ell > 3000$), which is not detected in our analysis due to the limited resolution of \emph{Planck} CMB maps.  Therefore, one should take extra care about the thermal SZ cluster residuals for higher resolution CMB maps.

\section*{Acknowledgements}

TC acknowledges the Overseas Research Scholarship from the School of Physics and Astronomy, The University of Manchester. MR and CD acknowledge support from an ERC Starting (Consolidator) Grant (no. 307209) and an STFC Consolidated Grant (ST/P000649/1). The authors thank Jens Chluba and Richard Battye for comments on the paper before submission, Naoki Itoh, and Ricardo G\'enova-Santos for comments after submission, and the anonymous referee for their useful comments on the paper.


\bibliographystyle{mn2e}
\bibliography{journals,lit} 




\bsp	
\label{lastpage}
\end{document}